\documentclass[11pt,oneside,reqno]{amsart}
\usepackage[latin9]{inputenc}
\usepackage[a4paper]{geometry}
\usepackage{color}
\usepackage{mathrsfs}
\usepackage{amstext}
\usepackage{amsthm}
\usepackage{amssymb}
\usepackage{graphicx}
\usepackage{setspace}
\makeatletter

\providecommand{\tabularnewline}{\\}

\numberwithin{equation}{section}
\numberwithin{figure}{section}
\numberwithin{table}{section}

\usepackage{amsfonts}
\usepackage{amstext}
\usepackage{amsthm}
\usepackage{multirow}
\usepackage{siunitx}
\usepackage{lscape}

\newtheorem*{asm}{Assumption}\newtheorem{thm}{Theorem}\newtheorem{lem}{Lemma}

\theoremstyle{definition}
\newtheorem{rem}{Remark}

\makeatother

\begin{document}
\title{GLS under monotone heteroskedasticity}
\author{Yoichi Arai}
\address{School of Social Sciences, Waseda University, 1-6-1 Nishiwaseda, Shinjuku-ku,
Tokyo 169-8050, Japan.}
\email{yarai@waseda.jp}
\author{Taisuke Otsu}
\address{Department of Economics, London School of Economics, Houghton Street,
London, WC2A 2AE, UK.}
\email{t.otsu@lse.ac.uk}
\author{Mengshan Xu}
\address{Department of Economics, University of Mannheim, L7 3-5, 68161, Mannheim,
Germany. }
\email{mengshan.xu@uni-mannheim.de}
\begin{abstract}
The generalized least square (GLS) is one of the most basic tools
in regression analyses. A major issue in implementing the GLS is estimation
of the conditional variance function of the error term, which typically
requires a restrictive functional form assumption for parametric estimation
or smoothing parameters for nonparametric estimation. In this paper,
we propose an alternative approach to estimate the conditional variance
function under nonparametric monotonicity constraints by utilizing
the isotonic regression method. Our GLS estimator is shown to be asymptotically
equivalent to the infeasible GLS estimator with knowledge of the conditional
error variance, and involves only some tuning to trim boundary observations,
not only for point estimation but also for interval estimation or
hypothesis testing. Our analysis extends the scope of the isotonic
regression method by showing that the isotonic estimates, possibly
with generated variables, can be employed as first stage estimates
to be plugged in for semiparametric objects. Simulation studies illustrate
excellent finite sample performances of the proposed method. As an
empirical example, we revisit Acemoglu and Restrepo's (2017) study
on the relationship between an aging population and economic growth
to illustrate how our GLS estimator effectively reduces estimation
errors.
\end{abstract}

\maketitle

\section{Introduction\label{sec:intro}}

The generalized least square (GLS) is one of the most basic tools
in regression analyses. It yields the best linear unbiased estimator
in the classical linear regression model, and has been studied extensively
in econometrics and statistics literature; see e.g., Wooldridge (2010,
Chapter 7) for a review. A major issue in implementing the GLS is
that the optimal weights given by the conditional error variance function
(say, $\sigma^{2}(\cdot)$) are typically unknown to researchers and
need to be estimated. One way to estimate $\sigma^{2}(\cdot)$ is
to specify its parametric functional form and estimate it by a parametric
regression for the squared OLS residuals of the original regression
on the specified covariates. However, economic theory rarely provides
exact functional forms of $\sigma^{2}(\cdot)$, and the feasible GLS
using misspecified $\sigma^{2}(\cdot)$ is no longer asymptotically
efficient (Cragg, 1983). To address this issue, Carroll (1982) and
Robinson (1987) proposed to estimate $\sigma^{2}(\cdot)$ nonparametrically
and established the asymptotic equivalence of the resulting feasible
GLS estimator with the infeasible one under certain regularity conditions.
This is a remarkable result, but it requires theoretically and practically
judicious choices of smoothing parameters, such as bandwidths, series
lengths, or numbers of neighbors. It should be noted that such smoothing
parameters appear in not only the point estimator but also its standard
error for inference, and their choices typically require some assumption
or knowledge of the smoothness of the conditional variance and associated
density functions, such as their differentiability orders.

In this paper, we propose an alternative approach to estimate the
conditional error variance function to implement the GLS by exploring
a shape constraint of $\sigma^{2}(\cdot)$ instead of its smoothness
as in Robinson (1987). As argued by Matzkin (1994), economic theory
often provides shape constraints for functions of economic variables,
such as monotonicity, concavity, or symmetry. In particular, we focus
on situations where $\sigma^{2}(\cdot)$ is known to be monotone in
its argument even though its exact functional form is unspecified,
and propose to estimate $\sigma^{2}(\cdot)$ by utilizing the method
of isotonic regression (see a review by Groeneboom and Jongbloed,
2014). It is known that the conventional isotonic regression estimator
typically yields piecewise constant function estimates and does not
involve any tuning parameters. Although the limiting behavior of the
isotonic regression estimator is less tractable (such as the $n^{1/3}$-consistency
and complicated limiting distribution), we show that our feasible
GLS estimator using the optimal weights by the isotonic estimator
with some trimming for boundary observations is asymptotically equivalent
to the infeasible GLS estimator. Furthermore, we can plug in this
isotonic estimator to estimate the asymptotic variance of the GLS
estimator for statistical inference.

For the linear model $Y=X^{\prime}\beta+U$ in the presence of heteroskedasticity
$\sigma^{2}(X)=E[U^{2}|X]$, using feasible GLS to improve the estimation
efficiency has a long history. On the one hand, several parametric
models have been proposed to estimate conditional error variance function
$\sigma^{2}(\cdot)$. See Remark \ref{rem:spec} below. On the other
hand, Carroll (1982) and Robinson (1987) estimated $\sigma^{2}(\cdot)$
with kernel and nearest neighbor estimator, respectively, and they
showed their semiparametric GLS estimators are asymptotically equivalent
to the infeasible GLS estimator and thus efficient. Compared to existing
parametric methods, our proposed method imposes monotonicity, a feature
implied by many parametric models, but it is nonparametric and does
not rely on any specific parametric function form.\footnote{Monotone heteroskedasticity is often observed in economic literature.
For example, Mincer (1974) argued that the variance of wages, when
conditioned on education, should increase with the level of education
because individuals with higher education have a broader array of
job choices. Ruud (2000) cited this argument and provided empirical
evidence in his Figure 18.1 based on the CPS data from March 1995.
Another example can be found in Example 8.6 of Wooldridge (2013, pp.
283-284), where he employed a univariate\textbf{ }conditional variance
function of log income to explain the heteroskedasticity observed
in net total financial wealth of people in the United States.} Compared to existing nonparametric methods, our proposed method involves
only some tuning to trim boundary observations which does not require
knowledge of the smoothness of the conditional variance and associated
density functions. In the Monte Carlo simulations, we show that our
proposed method outperforms the above-mentioned nonparametric methods
at almost every choice of smoothing parameters, while it performs
as well as parametric feasible GLS estimators with correctly specified
conditional error variance function.

The isotonic estimator can date back to the middle of the last century.
Earlier work includes Ayer \emph{et al.} (1955), Grenander (1956),
Rao (1969, 1970), and Barlow and Brunk (1972), among others. The isotonic
estimator of a regression function can be formulated as a least square
estimation with monotonicity constraints. Suppose that the conditional
expectation $E[Y|X]=m(X)$ is monotone increasing, for an iid random
sample $\{Y_{i},X_{i}\}_{i=1}^{n}$, the isotonic estimator is the
minimizer of the sum of squared errors, $\underset{m\in\mathcal{M}}{\min}\sum_{i=1}^{n}\{Y_{i}-m(X_{i})\}^{2},$
where $\mathcal{M}$ is the class of monotone increasing functions.
The minimizer can be calculated with the pool adjacent violators algorithm
(Barlow and Brunk, 1972), or equivalently by solving the greatest
convex minorant of the cumulative sum diagram $\{(0,0),(i,\sum_{j=1}^{i}Y_{j}),i=1,\ldots,n\}$,
where the corresponding $\{X_{i}\}_{i=1}^{n}$ are ordered sequence;
see Groeneboom and Jongbloed (2014) for a comprehensive discussion
of different aspects of isotonic regression. Moreover, recent developments
in the monotone single index model provide convenient and flexible
tools for combining monotonicity and multi-dimensional covariates.
In a monotone single index model, the conditional mean of $Y$ is
modeled as $E[Y|X]=m(X^{\prime}\alpha)$, and the monotone link function
$m(\cdot)$ is solved with isotonic regression. Balabdaoui, Durot
and Jankowski (2019) studied the monotone single index model with
the monotone least square method. Groeneboom and Hendrickx (2018),
Balabdaoui, Groeneboom and Hendrickx (2019), and Balabdaoui and Groeneboom
(2021) developed a score-type approach for the monotone single index
model. Their approach can estimate the single index parameter $\alpha$
and the link function $m(\cdot)$ at $n^{-1/2}$-rate and $n^{-1/3}$-rate
respectively. We employ their approach for the estimation of the conditional
variance function in the multivariate case. Recently, Babii and Kumar
(2023) applied the isotonic regression to their analysis of regression
discontinuity designs. To this end, Babii and Kumar (2023) extended
existing results concerning the boundary properties of Grenander's
estimator (e.g., those from Woodroofe and Sun, 1993, and Kulikov and
Lopuhaä, 2006) to derive the asymptotic distribution of their trimmed
isotonic regression discontinuity estimator. To regularize the isotonic
estimator in the weights of our proposed GLS estimator, we employ
a similar trimming strategy while adapting the theory of Babii and
Kumar (2023) to our context of the conditional variance estimation.
We contribute to this literature on isotonic regression by showing
that the isotonic estimates can be employed as first stage estimates
to be plugged in for semiparametric objects. Furthermore, we note
that our isotonic estimator involves generated variables (i.e., OLS
residuals), which make theoretical developments substantially different
from the existing ones.

This paper is organized as follows. In Section \ref{sec:case1}, we
consider the case where $\sigma^{2}(\cdot)$ is monotone in one covariate,
present our GLS estimator, and study its asymptotic properties. Section
\ref{sec:case2} extends our GLS approach to the case where $\sigma^{2}(\cdot)$
is specified by a monotone single index function. Section \ref{sec:numerical}
illustrates the proposed method by a simulation study and empirical
example.

\section{Heteroskedasticity by univariate covariate\label{sec:case1}}

We first consider the case where monotone heteroskedasticity is caused
by a single covariate. In particular, consider the following multiple
linear regression model 
\begin{equation}
Y=\alpha+\beta X+Z^{\prime}\gamma+U,\qquad E[U|X,Z]=0,\label{eq:model}
\end{equation}
where $X\in\mathscr{X}=[x_{L},x_{U}]$ is a scalar covariate with
compact support and $Z$ is a vector of other covariates. In this
section, we focus on the case where heteroskedasticity is caused by
the covariate $X$, i.e., 
\begin{equation}
E[U^{2}|X,Z]=E[U^{2}|X]=:\sigma^{2}(X),\label{eq:het}
\end{equation}
and $\sigma^{2}(\cdot)$ is a monotone increasing function. The case
of monotone decreasing $\sigma^{2}(\cdot)$ is analyzed analogously
(by setting $U^{2}$ as $-U^{2}$). In the setup (\ref{eq:het}),
we assume that the researcher knows which covariate should be included
in $\sigma^{2}(\cdot)$ based on economic theory or other prior information.
This setup should be considered as a useful benchmark to provide a
clear exposition of the main concept and the asymptotic properties
of the proposed monotone GLS estimator. Without the covariates $Z$,
the above model covers a bivariate regression model, and our approach
is new even in such a fundamental setup. Furthermore, this setup covers
the case where $X$ contained in (\ref{eq:het}) does not enter the
regression model (\ref{eq:model}) by setting $\beta=0$ (such a situation
is considered in our empirical illustration in Section \ref{sub:emp}).
Extensions to relax the assumption in (\ref{eq:het}) will be discussed
in Remark \ref{rem:discZ} and Section \ref{sec:case2}.

Let $\theta=(\alpha,\beta,\gamma^{\prime})^{\prime}$ be a vector
of the slope parameters and $W:=(1,X,Z^{\prime})^{\prime}$ so that
the model in (\ref{eq:model}) can be written as $Y=W^{\prime}\theta+U$.
Based on an iid sample $\{Y_{i},X_{i},Z_{i}\}_{i=1}^{n}$, the infeasible
GLS estimator for $\theta$ is written as 
\begin{equation}
\hat{\theta}_{\mathrm{IGLS}}=\left(\sum_{i=1}^{n}\sigma_{i}^{-2}W_{i}W_{i}^{\prime}\right)^{-1}\left(\sum_{i=1}^{n}\sigma_{i}^{-2}W_{i}Y_{i}\right),\label{eq:IGLS}
\end{equation}
where $\sigma_{i}^{2}=\sigma^{2}(X_{i})$. In order to make this estimator
feasible, various approaches have been proposed in the literature.

In this paper, we are concerned with the situation where the researcher
knows $\sigma^{2}(\cdot)$ is monotone in a particular regressor $X$
but its exact functional form is unspecified. In particular, by utilizing
knowledge of the monotonicity of $\sigma^{2}(\cdot)$, we propose
to estimate $\sigma^{2}(\cdot)$ by the isotonic regression from the
squared OLS residual on the regressor $X$. More precisely, let $\hat{\theta}_{\mathrm{OLS}}=\left(\sum_{i=1}^{n}W_{i}W_{i}^{\prime}\right)^{-1}\left(\sum_{i=1}^{n}W_{i}Y_{i}\right)$
be the OLS estimator for (\ref{eq:model}), and $\hat{U}_{j}=Y_{j}-W_{j}^{\prime}\hat{\theta}_{\mathrm{OLS}}$
be its residual. Then we estimate $\sigma^{2}(\cdot)$ by 
\begin{equation}
\hat{\sigma}^{2}(\cdot)=\text{isotonic regression function from }\{\hat{U}_{j}^{2}\}_{j=1}^{n}\text{ on }\{X_{j}\}_{j=1}^{n}.\label{eq:sigma}
\end{equation}

Although this estimator is shown to be consistent for $\sigma^{2}(\cdot)$
in the interior of support $[x_{L},x_{U}]$ of $X$, it is generally
biased at the lower boundary $x_{L}$, which may cause inconsistency
of the resulting GLS estimator. Therefore, we propose to trim observations
whose $X_{i}$'s are too close to $x_{L}$, and develop the following
feasible GLS estimator 
\begin{equation}
\hat{\theta}=\left(\sum_{i=1}^{n}\mathbb{I}\{X_{i}\ge q_{n}\}\hat{\sigma}_{i}^{-2}W_{i}W_{i}^{\prime}\right)^{-1}\left(\sum_{i=1}^{n}\mathbb{I}\{X_{i}\ge q_{n}\}\hat{\sigma}_{i}^{-2}W_{i}Y_{i}\right),\label{eq:t-FGLS}
\end{equation}
where $\mathbb{I}\{\cdot\}$ is the indicator function, and the trimming
term $q_{n}$ is set as the $(n^{-1/3})$-th sample quantile of $\{X_{i}\}_{i=1}^{n}$. 

Let $\mathcal{B}(a,R)$ be a ball around $a$ with radius $R$; for
$\varepsilon=U^{2}-\sigma^{2}(X)$, define $\sigma_{\varepsilon}^{2}(x)=E[\varepsilon^{2}|X=x]$.
To study the asymptotic properties of the proposed estimator $\hat{\theta}$,
we impose the following assumptions.

\begin{asm}$\quad$ 
\begin{description}
\item [{A1}] $\{Y_{i},X_{i},Z_{i}\}_{i=1}^{n}$ is an iid sample of $(Y,X,Z)$.
The support of $(X,Z)$ is convex with non-empty interiors and is
a subset of $\mathcal{B}(0,R)$ for some $R>0$. The support of $X$
is a compact interval $\mathscr{X}=[x_{L},x_{U}]$.
\item [{A2}] $\sigma^{2}:\mathscr{X}\to\mathbb{R}$ is a monotone increasing
function defined on $\mathscr{X}$, and $0<\sigma^{2}(x_{L})<\sigma^{2}(x_{U})<\infty$.
There exist positive constants $a_{0}$ and $M$ such that $E[|U|^{2s}|X=x]\le a_{0}s!M^{s-2}$
for all integers $s\ge2$ and $x\in\mathscr{X}$. For some positive
constant $\delta$, $\sigma^{2}(\cdot)$ is continuously differentiable
on $(x_{L},x_{L}+\delta)$, and $\sigma_{\varepsilon}^{2}(\cdot)$
is continuous on $(x_{L},x_{L}+\delta)$. 
\item [{A3}] $X$ has a continuous density function $f_{X}(\cdot)$ on
$\mathscr{X}$, and there exists a positive constant $b$ such that
$b<f_{X}(x)<\infty$ for all $x\in\mathscr{X}$. 
\end{description}
\end{asm}

Assumption A1 is standard. As pointed out in Balabdaoui, Groeneboom
and Hendrickx (2019, p.13), the compact support assumption can be
relaxed when $X$ follows a sub-Gaussian distribution. In this case,
the $L^{2}$-convergence rate of the isotonic estimator will decrease
from $O_{p}(n^{-1/3}\log n)$ to $O_{p}(n^{-1/3}(\log n)^{5/4})$.
Another impact of relaxing the distribution of $X$ (and $Z$) to
a sub-Gaussian one is on the concentration rate of $\max_{j}|\hat{U}_{j}^{2}-U_{j}^{2}|$
(see Appendix \ref{app:uni-proofs} for more details). This rate,
used in proving Lemma 1 and explaining the concentration of $T_{1}$
and $T_{2}$ in Appendix \ref{app-sub:thm_uni}, will inflate by a
factor of $\log n$. However, even with this change, we still have
$\max_{j}|\hat{U}_{j}^{2}-U_{j}^{2}|=o_{p}(n^{-1/3})$, which is the
key to show that the impact of substituting infeasible $U^{2}$ with
estimated $\hat{U}^{2}$ on isotonic estimators is asymptotically
negligible. Considering that the convergence rates of these aforementioned
terms are slowed down by a factor of $\log n$ at most, the validity
of the main results in this paper is preserved with sub-Gaussian covariates,
but the analytical derivation would become more cumbersome. For a
clearer and more concise exposition, we maintain the compact support
assumption on $X$. Assumption A2 is on the error term. The monotonicity
of $\sigma^{2}(\cdot)$ is the main assumption. The assumption on
arbitrary higher moments, which rules out some fat-tailed distributions,
is commonly used to obtain some maximal inequalities (cf. van der
Vaart and Wellner, 1996, Lemma 2.2.11, for a similar assumption).
Assumption A3 contains additional mild conditions on the density of
$X$.

We first present asymptotic properties of the conditional error variance
estimator $\hat{\sigma}^{2}(\cdot)$ in (\ref{eq:sigma}). Let $q_{n}^{*}$
be the $(n^{-1/3})$-th population quantile of $X$, $D_{A}^{L}[f](a)$
be the left derivative of the greatest convex minorant of a function
$f(\cdot)$ evaluated at $a\in A$, and $\{\mathcal{W}_{t}\}$ be
the standard Brownian motion. Also define $c^{*}=\lim_{n\to\infty}n^{1/3}(q_{n}^{*}-x_{L})$.
Assumption A3 guarantees $0<c^{*}<\infty$. Then we obtain the following
lemma for the behavior of $\hat{\sigma}^{2}(\cdot)$ around the boundary
$x_{L}$, which extends the result by Babii and Kumar (2023, Theorem
2.1(ii)) by allowing the generated variable $\hat{U}_{i}^{2}$ as
a regressand for $\hat{\sigma}^{2}(\cdot)$.

\begin{lem} \label{lem:trun_uni} Under Assumptions A1-A3 and $\lim_{x\downarrow x_{L}}\frac{d\sigma^{2}(x)}{dx}>0$,
it holds 
\begin{equation}
n^{1/3}\{\hat{\sigma}^{2}(q_{n})-\sigma^{2}(q_{n})\}\overset{d}{\to}D_{[0,\infty)}^{L}\left[\sqrt{\frac{\sigma_{\varepsilon}^{2}(x_{L})}{c^{*}f_{X}(x_{L})}}\mathcal{W}_{t}+\left(\lim_{x\downarrow x_{L}}\frac{d\sigma^{2}(x)}{dx}\right)c^{*}\left(\frac{1}{2}t^{2}-t\right)\right](1).\label{eq:consistency-2}
\end{equation}
\end{lem}

Based on this lemma, the asymptotic distribution of our feasible GLS
estimator $\hat{\theta}$ is obtained as follows.

\begin{thm} \label{thm:uni} Under Assumptions A1-A3, it holds
\[
\sqrt{n}(\hat{\theta}-\theta)\overset{d}{\to}N(0,E[\sigma^{-2}(X)WW^{\prime}]^{-1}),
\]
and the asymptotic variance matrix is consistently estimated by $\left(\frac{1}{n}\sum_{i=1}^{n}\hat{\sigma}_{i}^{-2}W_{i}W_{i}^{\prime}\right)^{-1}$.
\end{thm}

This theorem implies that our estimator $\hat{\theta}$ has the same
limiting distribution as the infeasible GLS estimator $\hat{\theta}_{\mathrm{IGLS}}$
and thus achieves the semiparametric efficiency bound. This result
extends the scope of the isotonic regression method by showing that
the isotonic estimates, possibly with generated variables, can be
employed as first stage estimates to be plugged in for semiparametric
objects. We re-emphasize that $\hat{\theta}$ involves only a trimming
term $q_{n}$, the $(n^{-1/3})$-th sample quantile of $\{X_{i}\}_{i=1}^{n}$.\footnote{Although our estimator $\hat{\theta}$ in (\ref{eq:t-FGLS}) does
not involve any tuning constant, the trimming term $q_{n}$ should
be understood as the $c\cdot(n^{-1/3})$-th sample quantile of $\{X_{i}\}_{i=1}^{n}$,
where the tuning constant is set as $c=1$. Indeed Theorem \ref{thm:uni}
holds true with any $c>0$. If we compare with other nonparametric
methods, smoothing parameters, such as bandwidths, series lengths,
and neighbors, typically require two constants to implement. For example,
for the bandwidth parameter $b=c_{1}n^{-c_{2}}$, researchers need
to choose $c_{1}$ and $c_{2}$. The constant $c_{1}$, which is analogous
to $c$ above, can be any positive number. However, they also need
to choose a positive constant $c_{2}$ whose upper bound typically
depends on (unknown) smoothness of underlying functions.}

\begin{rem}\label{rem:discZ} {[}Extensions of (\ref{eq:het}){]}
The benchmark setup $E[U^{2}|X,Z]=\sigma^{2}(X)$ considered in this
section can be extended in various ways. First, an extension to a
single index model (say, $E[U^{2}|X,Z]=\sigma^{2}(X\eta_{x}+Z^{\prime}\eta_{z})$)
will be discussed in the next section. Second, the model in (\ref{eq:model})-(\ref{eq:het})
can be extended to the case where the conditional variance varies
with discrete covariates $Z$ (or its subvector), say $E[U^{2}|X,Z=z]=\sigma_{z}^{2}(X)$
with monotone functions $\sigma_{z}^{2}(\cdot)$ for $z\in\{z^{(1)},\ldots,z^{(D)}\}$.
In this case, we can implement the isotonic regression for each group
categorized by $z$, and construct the feasible GLS estimator in an
analogous way as (\ref{eq:t-FGLS}). Third, our approach may be extended
to the additive monotone heteroskedasticity, say $E[U^{2}|X,Z]=\sigma_{x}^{2}(X)+\sigma_{z}^{2}(Z)$
with monotone functions $\sigma_{x}^{2}(\cdot)$ and $\sigma_{z}^{2}(\cdot)$.
Although formal analysis is beyond the scope of this paper, the results
in Mammen and Yu (2007) suggest that the isotonic estimators for additive
functions converge at similar rates as the univariate case, and we
conjecture that a similar result as Theorem \ref{thm:uni} can be
obtained. Finally, when the conditional error variance function is
multiplicative, say $E[U^{2}|X,Z]=\sigma_{x}^{2}(X)\sigma_{z}^{2}(Z)$,
and the researcher knows the form of $\sigma_{z}^{2}(\cdot)$ (e.g.,
$Z$ is household size and $\sigma_{z}^{2}(Z)=Z^{2}$), then our feasible
GLS estimator can be applied to observations reweighted by $1/\sigma_{z}(Z)$.
\end{rem}

\begin{rem}\label{rem:thm1} {[}Monotonicity testing{]} Monotonicity
is an assumption that can be tested. For observable random variables
$(Y,X)$, several methods have been developed to test whether $E[Y|X]$
is monotone increasing in $X$; see, e.g., Ghosal, Sen and van der
Vaart (2000), Hall and Heckman (2000), Dümbgen and Spokoiny (2001),
Chetverikov (2019), and Hsu, Liu and Shi (2019), among others. All
these tests can be adapted for our case, testing the monotonicity
of $\sigma^{2}(\cdot)$ with generated $\{\hat{U}_{j}^{2}\}_{j=1}^{n}$
and observed $\{X_{j}\}_{j=1}^{n}$. Since Assumptions A1-A2 and $\hat{\theta}_{\mathrm{OLS}}-\theta=O_{p}(n^{-1/2})$
imply $\hat{U}_{j}^{2}-U_{j}^{2}=O_{p}(n^{-1/2}\log n)$ uniformly
over $j=1,\ldots,n$, the critical values of these tests can be adjusted
accordingly to maintain a proper asymptotic size. \end{rem}

\begin{rem}\label{rem:thm1_non_mono} {[}Misspecification of $E[U^{2}|X,Z]${]}
We want to note that even if the assumption in (\ref{eq:het}) is
violated (e.g., $E[U^{2}|X,Z]$ varies with $Z$ or $E[U^{2}|X,Z]=\sigma^{2}(X)$
with non-monotone $\sigma^{2}(\cdot)$), our feasible GLS estimator
$\hat{\theta}$ in (\ref{eq:t-FGLS}) is still consistent for $\theta$
due to $E[U|X,Z]=0$, and asymptotically normal at the $\sqrt{n}$-rate
with the limiting distribution
\[
\sqrt{n}(\hat{\theta}-\theta)\overset{d}{\to}N(0,E[\rho(X)^{-1}WW^{\prime}]^{-1}E[\rho(X)^{-2}E[U^{2}|X,Z]WW^{\prime}]E[\rho(X)^{-1}WW^{\prime}]^{-1}),
\]
where $\rho(\cdot)=\arg\min_{m\in\mathcal{M}}E[\{U^{2}-m(X)\}^{2}]$
for the class of monotone increasing functions $\mathcal{M}$. Since
$\hat{\sigma}^{2}(\cdot)$ can estimate $\rho(\cdot)$, then the asymptotic
variance matrix can be consistently estimated by
\begin{equation}
\left(\frac{1}{n}\sum_{i=1}^{n}\hat{\sigma}^{-2}(X_{i})W_{i}W_{i}^{\prime}\right)^{-1}\left(\frac{1}{n}\sum_{i=1}^{n}\hat{\sigma}^{-4}(X_{i})\hat{U}_{i}^{2}W_{i}W_{i}^{\prime}\right)\left(\frac{1}{n}\sum_{i=1}^{n}\hat{\sigma}^{-2}(X_{i})W_{i}W_{i}^{\prime}\right)^{-1}.\label{eq:sand}
\end{equation}
This misspecification robust variance estimator is analogous to the
one proposed by Cragg (1992) for the feasible GLS estimator with parametrically
specified models for the conditional error variance $E[U^{2}|X,Z]$.\footnote{Based on simulation studies, Cragg (1992) recommended to use his misspecification
robust variance estimator even when the parametric form of heteroskedasticity
is correctly specified. Although a similar analysis is beyond the
scope of this paper, we also recommend to employ the variance estimator
(\ref{eq:sand}) in practice due to its consistency regardless of
the assumption in (\ref{eq:het}).} \end{rem}

\begin{rem}\label{rem:IV} {[}Endogenous regressor{]} The result
of Theorem \ref{thm:uni} can also be extended to some linear instrumental
variable (IV) regression model. For notational simplicity, consider
the following univariate IV regression:
\[
Y=\alpha+\beta X+U,\qquad E[U|Z]=0,
\]
where $X$ is a scalar endogenous regressor and $Z$ is a scalar IV,
and we further assume $E[X|Z]=\eta+\gamma Z$ for some parameters
$(\eta,\gamma)$. This linearity assumption on $E[X|Z]$ is not essential,
and may be relaxed by some nonparametric estimator of $E[X|Z]$. In
this setup, the optimal instrument for estimating $(\alpha,\beta)^{\prime}$
is given by (see, e.g., Newey, 1993)
\[
E\left[\frac{\partial(Y-\alpha-\beta X)}{\partial(\alpha,\beta)^{\prime}}\bigg|Z\right]E[U^{2}|Z]^{-1}=-\left(\begin{array}{cc}
1 & 0\\
\eta & \gamma
\end{array}\right)\left(\begin{array}{c}
1\\
Z
\end{array}\right)v^{-2}(Z),
\]
where $v^{2}(\cdot)=E[U^{2}|Z=\cdot]$. Under the assumption of $\gamma\ne0$
(i.e., the IV is relevant), the optimal IV estimator is obtained by
the method of moments estimator of the following moment condition:
\begin{equation}
E\left[\left(\begin{array}{c}
1\\
Z
\end{array}\right)v^{-2}(Z)(Y-\alpha-\beta X)\right]=0.\label{eq:IV}
\end{equation}
Under the monotonicity assumption of $v^{2}(\cdot)$, we can obtain
the isotonic estimator $\hat{v}^{2}(\cdot)$ for $v^{2}(\cdot)$ by
regressing the squared residuals $\hat{e}^{2}=(Y-\tilde{\alpha}-\tilde{\beta}X)^{2}$
for an initial estimator $(\tilde{\alpha},\tilde{\beta})$ (e.g.,
the two-stage least squares estimator) on $Z$. The resulting estimator,
$\hat{v}^{2}(\cdot)$, should have the same properties as those of
$\hat{\sigma}^{2}(\cdot)$ presented in Lemma 1, where $q_{n}$ is
replaced with the $(n^{-1/3})$-th sample quantile of $\{Z_{i}\}_{i=1}^{n}$.
Based on this isotonic estimator, a feasible optimal IV estimator
$\hat{\theta}_{\mathrm{IV}}=(\hat{\alpha}_{\mathrm{IV}},\hat{\beta}_{\mathrm{IV}})^{\prime}$
is given by
\[
\hat{\theta}_{\mathrm{IV}}=\left(\sum_{i=1}^{n}\mathbb{I}\{Z_{i}\ge q_{n}\}\hat{v}^{-2}(Z_{i})\left(\begin{array}{c}
1\\
Z_{i}
\end{array}\right)(1,X_{i})\right)^{-1}\left(\sum_{i=1}^{n}\mathbb{I}\{Z_{i}\ge q_{n}\}\hat{v}^{-2}(Z_{i})\left(\begin{array}{c}
1\\
Z_{i}
\end{array}\right)Y_{i}\right).
\]
By applying the same arguments for Theorem \ref{thm:uni}, we can
show that $\hat{\theta}_{\mathrm{IV}}$ is asymptotically equivalent
to the infeasible optimal IV estimator based on (\ref{eq:IV}) with
known $v^{2}(\cdot)$. \end{rem}

\section{Heteroskedasticity by multivariate covariates\label{sec:case2}}

We now consider the model
\begin{equation}
Y=\alpha+X^{\prime}\beta+Z^{\prime}\gamma+U,\qquad E[U|X,Z]=0,\label{eq:index}
\end{equation}
where $X$ is a vector of covariates. This section focuses on the
case where heteroskedasticity takes the form of a monotone single
index function of $X$ with unknown parameters $\eta_{0}$, i.e.,
$E[U^{2}|X,Z]=E[U^{2}|X]=\sigma^{2}(X^{\prime}\eta_{0})$ for a monotone
increasing function $\sigma^{2}(\cdot)$. Single index models are
known to be more flexible than parametric models and achieve dimension
reduction relative to nonparametric models. 

\begin{rem}\label{rem:spec} First, the monotone index model $\sigma^{2}(X^{\prime}\eta_{0})$
covers several existing parametric models. Popular examples include
$\sigma^{2}(X)=C(X^{\prime}\eta_{0})^{2-2\lambda}$ (Box and Hill,
1974), $\sigma^{2}(X)=C\exp(\lambda(X^{\prime}\eta_{0}))$ (Bickel,
1978), $\sigma^{2}(X)=C\{1+\lambda(X^{\prime}\eta_{0})^{2}\}$ (Fuller,
1980) for some constants $C>0$ and $\lambda$; interestingly, all
these parametric functions are monotone increasing (or decreasing)
in the index of $X$. Second, although the setup $E[U^{2}|X,Z]=\sigma^{2}(X^{\prime}\eta_{0})$
assumes that the researcher knows which (sub-)vector of covariates
should be included in $\sigma^{2}(\cdot)$, researchers do \emph{not}
have to select those covariates in the case where such prior information
is unavailable. They can simply re-define the model in (\ref{eq:index})
without covariates $Z$ (or equivalently specify as $E[U^{2}|X,Z]=\sigma^{2}(X^{\prime}\eta_{0}+Z^{\prime}\eta_{z0})$).
Our asymptotic theory below applies even if some covariates are irrelevant
for $E[U^{2}|X,Z]$. \end{rem}

For identification, $\eta_{0}$ is normalized as $||\eta_{0}||=1$.
Define 
\begin{equation}
\sigma_{\eta}^{2}(a)=E[\sigma^{2}(X^{\prime}\eta_{0})|X^{\prime}\eta=a].\label{eq:multi_mean}
\end{equation}
We show in Lemma \ref{lem:BGH_e_hat} that $\sigma{}^{2}(\cdot)$
and $\eta_{0}$ can be consistently estimated by extending the method
proposed in Balabdaoui, Groeneboom and Hendrickx (2019) (BGH hereafter)
and Balabdaoui and Groeneboom (2021) to allow generated variables.
In particular, for a given $\eta$, define the isotonic regression
of $\{\hat{U}_{i}^{2}\}_{i=1}^{n}$ on $\{X_{i}^{\prime}\eta\}_{i=1}^{n}$
as 
\begin{equation}
\hat{\sigma}_{\eta}^{2}=\arg\min_{m\in\mathcal{M}}\frac{1}{n}\sum_{i=1}^{n}\{\hat{U}_{i}^{2}-m(X_{i}^{\prime}\eta)\}^{2},\label{eq:p_gamma}
\end{equation}
where $\mathcal{M}$ is the set of monotone increasing functions defined
on $\mathbb{R}$. Based on this, $\hat{\eta}$ can be estimated by
minimizing the square sum of a score function. For example, the simple
score estimator in the spirit of BGH and Balabdaoui and Groeneboom
(2021) is given by 
\begin{equation}
\hat{\eta}=\text{arg}\min_{\eta}\left\Vert \frac{1}{n}\sum_{i=1}^{n}X_{i}\{\hat{U}_{i}^{2}-\hat{\sigma}_{\eta}^{2}(X_{i}^{\prime}\eta)\}\right\Vert ^{2},\label{eq:simplescore}
\end{equation}
where $\left\Vert \cdot\right\Vert $ is the Euclidean norm: $\left\Vert a\right\Vert =\sqrt{\sum_{j=1}^{k}a_{j}^{2}}$
for $a=(a_{1},\dots,a_{k})^{\prime}\in\mathbb{R}^{k}$.

Letting $\hat{\sigma}_{i}^{2}=\hat{\sigma}_{\hat{\eta}}^{2}(X_{i}^{\prime}\hat{\eta})$
and $W=(1,X^{\prime},Z^{\prime})^{\prime}$, we propose the following
GLS estimator 
\begin{equation}
\hat{\theta}=\left(\sum_{i=1}^{n}\mathbb{I}\{X_{i}^{\prime}\hat{\eta}\ge q_{n}\}\hat{\sigma}_{i}^{-2}W_{i}W_{i}^{\prime}\right)^{-1}\left(\sum_{i=1}^{n}\mathbb{I}\{X_{i}^{\prime}\hat{\eta}\ge q_{n}\}\hat{\sigma}_{i}^{-2}W_{i}Y_{i}\right),\label{eq:est-multi}
\end{equation}
where $q_{n}$ is the $(n^{-1/3})$-th sample quantile of $\{X_{i}^{\prime}\hat{\eta}\}_{i=1}^{n}$. 

To avoid unnecessarily heavy notations, in the multivariate case,
we redefine some notations, which have similar meanings to those used
in Section \ref{sec:case1}. Define $\varepsilon=U^{2}-\sigma^{2}(X^{\prime}\eta_{0})$,
$\sigma_{\varepsilon}^{2}(\cdot)=E[\varepsilon^{2}|X^{\prime}\eta_{0}=\cdot]$,
$x_{L}=\inf_{x\in\mathscr{X}}(x^{\prime}\eta_{0})$, and $x_{U}=\sup_{x\in\mathscr{X}}(x^{\prime}\eta_{0})$.
Let $f_{X}(\cdot)$ be the density function of the random variable
$X^{\prime}\eta_{0}$. Let $q_{n}^{*}$ be the $(n^{-1/3})$-th population
quantile of $X^{\prime}\eta_{0}$, $q_{n}$ be the $(n^{-1/3})$-th
sample quantile of $\{X_{i}^{\prime}\hat{\eta}\}_{i=1}^{n}$, $c^{*}=\lim_{n\to\infty}n^{1/3}(q_{n}^{*}-x_{L})$,
and $D_{A}^{L}[f](a)$ be the left derivative of the greatest convex
minorant of function $f(\cdot)$ evaluated at $a\in A$. Let $\text{dim}(w)$
be the dimension of a vector $w$. 

\begin{asm}$\quad$ 
\begin{description}
\item [{M1}] $\{Y_{i},X_{i},Z_{i}\}_{i=1}^{n}$ is an iid sample of $(Y,X,Z)$.
The support of $(X,Z)$, $\mathscr{X}\times\mathscr{Z}$, is convex
with non-empty interiors and is a subset of $\mathcal{B}(0,R)$ for
some $R>0$. 
\item [{M2}] (i) There exists $\delta_{0}>0$ such that the function $a\mapsto\sigma_{\eta}^{2}(a)$
defined in (\ref{eq:multi_mean}) is monotone increasing on $I_{\eta}=\{x^{\prime}\eta,x\in\mathscr{X}\}$
for each $\eta\in\mathcal{B}(\eta_{0},\delta_{0})$. (ii) $0<\inf_{a\in I_{\eta}}\sigma_{\eta}^{2}(a)<\sup_{a\in I_{\eta}}\sigma_{\eta}^{2}(a)<\infty$
for each $\eta\in\mathcal{B}(\eta_{0},\delta_{0})$. (iii) There exist
positive constants $a_{0}$ and $M$ such that $E[|U|^{2s}|X=x]\le a_{0}s!M^{s-2}$
for all integers $s\ge2$ and $x\in\mathscr{X}$. (iv) $\sigma_{\eta}^{2}(\cdot)$
is continuously differentiable on $I_{\eta}$ for each $\eta\in\mathcal{B}(\eta_{0},\delta_{0})$.
(v) $\sigma_{\varepsilon}^{2}(\cdot)$ is continuous on $(x_{L},x_{L}+\delta_{1})$
for some $\delta_{1}>0$. 
\item [{M3}] The random variable $X^{\prime}\eta_{0}$ has a density function
$f_{X}(\cdot)$ that is continuous on $I_{\eta_{0}}$. There exists
some real positive numbers $\underline{b}$ and $\overline{b}$, such
that $0<\underline{b}<f_{X}(a)<\overline{b}<\infty$ holds for all
$a\in$$I_{\eta_{0}}$. 
\item [{M4}] For each $\eta\in\mathcal{B}(\eta_{0},\delta_{0})$, the mapping
$a\mapsto E[X|X^{\prime}\eta=a]$ defined on $I_{\eta}$ is boun\textcolor{black}{ded
and has a finite total variation.} 
\item [{M5}] $\mathrm{Cov}[X^{\prime}(\eta_{0}-\eta),\text{ }\sigma^{2}(X^{\prime}\eta_{0})|X^{\prime}\eta]\neq0$
almost surely for each $\eta\ne\eta_{0}$. 
\item [{M6}] $B:=\int(x-E[X|x^{\prime}\eta_{0}])(x-E[X|x^{\prime}\eta_{0}])^{\prime}\left.\frac{d\sigma^{2}(a)}{da}\right|_{a=x^{\prime}\eta_{0}}dP(x)$
has rank $\text{dim}(\eta_{0})-$1. 
\end{description}
\end{asm}

Assumptions M1-M3 are analogs of Assumptions A1-A3, respectively.
The main assumption is the monotonicity of $\sigma_{\eta}^{2}(\cdot)$.
Assumptions M4-M6 are additional regularity conditions for the monotone
index model. By Assumption M1, we have $-\infty<x_{L}<x_{U}<\infty$.
Then similar to Lemma \ref{lem:trun_uni}, we obtain the following
lemma for the behavior of $\hat{\sigma}_{\hat{\eta}}^{2}(\cdot)$
around $x_{L}$.

\begin{lem} \label{lem:trun_multi} Under Assumptions M1-M6 and $\lim_{a\downarrow x_{L}}\frac{d\sigma^{2}(a)}{da}>0$,
it holds 
\[
n^{1/3}\{\hat{\sigma}_{\hat{\eta}}^{2}(q_{n})-\sigma^{2}(q_{n})\}\overset{d}{\to}D_{[0,\infty)}^{L}\left[\sqrt{\frac{\sigma_{\varepsilon}^{2}(x_{L})}{c^{*}f_{X}(x_{L})}}\mathcal{W}_{t}+\left(\lim_{a\downarrow x_{L}}\frac{d\sigma^{2}(a)}{da}\right)c^{*}\left(\frac{1}{2}t^{2}-t\right)\right](1).
\]
\end{lem}

Based on this lemma, the asymptotic distribution of the GLS estimator
$\hat{\theta}$ in (\ref{eq:est-multi}) is obtained as follows. Let
$\sigma_{i}^{2}=\sigma^{2}(X_{i}^{\prime}\eta_{0})$.

\begin{thm} \label{thm:multi}Under Assumptions M1-M6, it holds 
\[
\sqrt{n}(\hat{\theta}-\theta)\overset{d}{\to}N(0,E[\sigma^{-2}(X^{\prime}\eta_{0})WW^{\prime}]^{-1}),
\]
and the asymptotic variance matrix is consistently estimated by $\left(\frac{1}{n}\sum_{i=1}^{n}\hat{\sigma}_{i}^{-2}W_{i}W_{i}^{\prime}\right)^{-1}$.
\end{thm}

Similar comments to Theorem \ref{thm:uni} apply here. Our estimator
$\hat{\theta}$ is asymptotically equivalent to the infeasible GLS
estimator $\hat{\theta}_{\mathrm{IGLS}}$. In terms of technical contribution,
our theoretical analysis generalizes existing ones in, e.g., Babii
and Kumar (2023), BGH, and Balabdaoui and Groeneboom (2021) to accommodate
generated variables. Similar to Remark \ref{rem:thm1_non_mono}, even
when the monotonicity assumption of $\sigma_{\eta}^{2}(\cdot)$ is
violated, $\hat{\theta}$ is still consistent for $\theta$ and asymptotically
normal at the $\sqrt{n}$-rate with certain robust asymptotic variance.
Furthermore, endogenous regressors can be accommodated as in Remark
\ref{rem:IV}.

\begin{rem} We can suggest two informal robustness checks for the
monotone index assumption in (\ref{eq:multi_mean}). One is to compute
the standard errors robust to possible misspecification obtained in
the same manner as Remark \ref{rem:thm1_non_mono} and compare them
to those in Theorem \ref{thm:multi}. This can serve as a robustness
check for the monotone specification given variables of the conditional
error variance functions. Another is to report the results for the
specification where all exogenous variables are included to $\sigma^{2}(\cdot)$
in addition to those for the chosen specifications. A large difference
between these results can be a sign of the misspecification of the
chosen ones. See Section \ref{sub:emp} for illustration. \end{rem}

\begin{rem}\label{rem:spec1} In this section, we employ the monotone
single index structure to model the multivariate conditional variance
function. This strategy allows us to strike a balance between robustness
and mitigating the curse of dimensionality. Indeed, the current specification
can be extended to the multiple index model $E[U^{2}|X=x]=x_{0}^{\prime}\eta_{0}+\sum_{i=1}^{M}G_{i}(x_{i}^{\prime}\eta_{i})$,
for $X=(X_{0}^{\prime},X_{1}^{\prime}\dots,X_{M}^{\prime})^{\prime}$,
where $\{G_{i}(\cdot)\}_{i=1}^{M}$ are unknown monotone increasing
functions. For the case of $M=1$, this model simplifies to a monotone
partially linear single index model whose properties have been studied
by Xu and Otsu (2020). We are optimistic that, under certain regularity
conditions, similar results as in this section can be obtained. To
the best of our knowledge, we have not come across any works that
discuss the multiple monotone index model with $M>1$ even for the
conventional regression setup for $E[Y|X=x]$. A possible solution
could be derived by combining the existing literature on the monotone
single index model (as cited in Section \ref{sec:intro}) with the
literature on the monotone additive model (for instance, Mammen and
Yu, 2007). Another potential extension involves employing the nonparametric
framework of Fang, Guntuboyina and Sen (2021) to model the multivariate
conditional variance function. This framework is free of parametric
structure, and it requires the true conditional variance to be entirely
monotone increasing in its arguments, i.e., $\sigma^{2}(x_{1},z_{1})\leq\sigma^{2}(x_{2},z_{2})$
if only if $x_{1}\leq x_{2}$ and $z_{1}\leq z_{2}$. Explorations
of these extensions exceed the scope of this paper, and we leave them
for future research. \end{rem}

\section{Numerical illustrations\label{sec:numerical}}

\subsection{Simulation\label{sub:sim}}

We now investigate the finite sample properties of the proposed GLS
estimator by a Monte Carlo experiment. We follow the simulation design
by Cragg (1983) and Newey (1993). The first data generating process,
denoted by DGP1, is the heteroskedastic linear model with a univariate
covariate and normally distributed disturbance:\footnote{Normal random variables are not compactly supported, and hence it
violates Assumption A1. However, as discussed in the remark on Assumption
A1, this assumption can be relaxed.} 
\begin{eqnarray}
Y_{i} & = & \beta_{0}+\beta_{1}X_{i}+u_{i},\qquad u_{i}=\sigma_{i}\varepsilon_{i},\qquad\varepsilon_{i}\sim N(0,1),\nonumber \\
\beta_{0} & = & \beta_{1}=1,\qquad\log(X_{i})\sim N(0,1),\qquad X_{i}\mbox{ and }\varepsilon_{i}\mbox{ are independent},\nonumber \\
\sigma_{i}^{2} & = & .1+.2X_{i}+.3X_{i}^{2}.\label{eq:DGP1}
\end{eqnarray}
We consider three sample sizes, $n=50$, 100, and 500. The number
of replications is set to 1,000.

In addition to the feasible GLS estimator with monotone heteroskedasticity
(MGLS), we consider the ordinary least squares (OLS), infeasible generalized
least squares (GLS), feasible GLS (FGLS), and nearest neighbor estimators
(k-NN). GLS requires knowledge of the conditional error variance function
(\ref{eq:DGP1}), including the values of the coefficients. In contrast,
FGLS proceeds with the known functional form, but the coefficients
are estimated. The ``k-NN automatic'' chooses the number of neighbors
by a cross-validation procedure suggested by Newey (1993). All the
estimators except OLS are the weighted least squares estimators, and
their differences come from how the weights are calculated. Following
Newey (1993), we calculate the weights for each method by taking a
ratio of the predicted squared residual to the estimated variance
of the disturbance, censoring the result below $0.04$.

Table \ref{tab:univariate} presents the simulation results for estimation.
The first column shows the estimation methods, and the following two
columns show the root mean-squared error (RMSE) and mean absolute
error (MAE) for DGP1 with $n=50$. The results for GLS report the
levels of the RMSE and MAE, and those for others are their ratios
relative to GLS. The next two columns give the corresponding results
with $n=100$ and the last two columns with $n=500$. Two rows for
each estimator show the results for $\beta_{0}$ and $\beta_{1}$,
respectively. The inefficiency and inaccuracy of OLS are apparent.
FGLS performs quite well, and this is natural when the conditional
error variance functions are correctly specified. The performance
of k-NN varies with the choice of $k$ and is in between OLS and FGLS.
We observe that the performance of MGLS is better than k-NN in every
choice of smoothing parameters. The result of MGLS is comparable to
that of FGLS if not better. MGLS's independence of a smoothing parameter
is clearly desirable. We also note that MGLS performs well even for
$n=50$.

The last four columns of Table \ref{tab:univariate} present the results
for DGP2 with a homoskedastic error: 
\begin{eqnarray*}
Y_{i} & = & \beta_{0}+\beta_{1}X_{i}+u_{i},\qquad u_{i}\sim N(0,1),\\
\beta_{0} & = & \beta_{1}=1,\qquad\log(X_{i})\sim N(0,1),\qquad X_{i}\mbox{ and }u_{i}\mbox{ are independent}.
\end{eqnarray*}
For DGP2, all estimators work reasonably well although the performance
of k-NN with $k=6$ is worse than others.

\begin{landscape} 
\begin{table}[htp]
{\small{}\caption{{\small{}\label{tab:univariate}}Simulation: Estimation with univariate
covariate}
}{\small\par}
\begin{raggedright}
\begin{tabular}{|l|llcccc|cccccc|}
\hline 
 & \multicolumn{6}{c|}{DGP1} & \multicolumn{6}{c|}{DGP2}\tabularnewline
\hline 
 & \multicolumn{2}{c}{$n=50$} & \multicolumn{2}{c}{$n=100$} & \multicolumn{2}{c|}{$n=500$} & \multicolumn{2}{c}{$n=50$} & \multicolumn{2}{c}{$n=100$} & \multicolumn{2}{c|}{$n=500$}\tabularnewline
Estimator  & RMSE & MAE & RMSE & MAE & RMSE & MAE & RMSE & MAE & RMSE & MAE & RMSE & MAE\tabularnewline
\hline 
GLS (infeasible) & 0.132 & 0.085 & 0.093  & 0.059  & 0.041  & 0.028  & 0.194 & 0.122 & 0.133  & 0.088  & 0.057  & 0.039 \tabularnewline
 & 0.157 & 0.100 & 0.108  & 0.073  & 0.048  & 0.032  & 0.083 & 0.046 & 0.055  & 0.034  & 0.021  & 0.014 \tabularnewline
OLS  & 3.103 & 2.856 & 3.831  & 3.479  & 5.574  & 4.495  & 1.000 & 1.000 & 1.000  & 1.000  & 1.000  & 1.000 \tabularnewline
 & 2.072 & 2.098 & 2.543  & 2.370  & 3.377  & 2.971  & 1.000 & 1.000 & 1.000  & 1.000  & 1.000  & 1.000 \tabularnewline
FGLS  & 1.279 & 1.210 & 1.245  & 1.233  & 1.598  & 1.152  & 1.032 & 1.041 & 1.026  & 1.033  & 1.024  & 1.067 \tabularnewline
 & 1.427 & 1.268 & 1.406  & 1.280  & 1.271  & 1.242  & 1.090 & 1.092 & 1.075  & 1.036  & 1.088  & 1.090 \tabularnewline
k-NN (Automatic) & 1.630 & 1.373 & 1.633  & 1.511  & 1.355  & 1.167  & 1.123 & 1.081 & 1.130  & 1.081  & 1.181  & 1.138 \tabularnewline
 & 1.535 & 1.427 & 1.606  & 1.431  & 1.424  & 1.267  & 1.092 & 1.074 & 1.065  & 1.006  & 1.197  & 1.097 \tabularnewline
k-NN ($k=6$)  & 1.554 & 1.361 & 1.525  & 1.498  & 1.474  & 1.417  & 1.274 & 1.243 & 1.253  & 1.155  & 1.359  & 1.276 \tabularnewline
 & 1.466 & 1.421 & 1.472  & 1.462  & 1.454  & 1.459  & 1.178 & 1.143 & 1.177  & 1.114  & 1.350  & 1.344 \tabularnewline
k-NN ($k=15$)  & 1.600 & 1.386 & 1.566  & 1.365  & 1.251  & 1.108  & 1.037 & 1.076 & 1.079  & 1.059  & 1.081  & 1.140 \tabularnewline
 & 1.520 & 1.398 & 1.546  & 1.408  & 1.247  & 1.197  & 1.003 & 1.046 & 1.037  & 1.012  & 1.066  & 1.053 \tabularnewline
k-NN ($k=24$)  & 1.781 & 1.568 & 1.685  & 1.457  & 1.291  & 1.160  & 1.011 & 1.039 & 1.039  & 0.980  & 1.044  & 1.098 \tabularnewline
 & 1.630 & 1.560 & 1.673  & 1.471  & 1.312  & 1.246  & 1.002 & 1.026 & 1.015  & 0.994  & 1.038  & 1.025 \tabularnewline
MGLS  & 1.379 & 1.285 & 1.326  & 1.279  & 1.113  & 1.129  & 1.039 & 1.091 & 1.049  & 1.075  & 1.027  & 1.075 \tabularnewline
 & 1.327 & 1.214 & 1.332  & 1.249  & 1.113  & 1.144  & 1.043 & 1.051 & 1.051  & 1.058  & 1.055  & 1.066 \tabularnewline
\hline 
\end{tabular}
\par\end{raggedright}
\raggedright{}\vspace{0.5cm}
{\footnotesize{}Note: ``RMSE'' and ``MAE'' stand for the root mean
squared error and mean absolute error, respectively. The results for
GLS report}\\
{\footnotesize{}the levels of the RMSE and MAE, and those for others
are their ratios relative to GLS.}{\footnotesize\par}
\end{table}
 \end{landscape}

Next, we consider the heteroskedastic linear models with multivariate
covariates, denoted by DGP3: 
\begin{eqnarray}
Y_{i} & = & \beta_{0}+\beta_{1}X_{1i}+\beta_{2}X_{2i}+u_{i},\qquad u_{i}=\sigma_{i}\varepsilon_{i},\qquad\varepsilon_{i}\sim N(0,1),\nonumber \\
\beta_{0} & = & \beta_{1}=\beta_{2}=1,\qquad\log(X_{1i}),\ \log(X_{2i})\sim N(0,1),\qquad X_{1i},\ X_{2i}\mbox{ and }\varepsilon_{i}\mbox{ are independent},\nonumber \\
\sigma_{i}^{2} & = & .2(X_{1i}+X_{2i})^{2}.\label{eq:DGP3}
\end{eqnarray}
The conditional error variance function of DPG3 is of a monotone single
index structure. Using the notation in (\ref{eq:multi_mean}), DGP3
corresponds to the structure with $\sigma^{2}(a)=a^{2}$, $X^{\prime}=(X_{1},X_{2})$,
and $\eta_{0}=(\sqrt{.2},\sqrt{.2})^{\prime}$. The left panel of
Table \ref{tab:multivariate} shows the results of DGP3 in the same
manner as Table \ref{tab:univariate}. For each estimation method,
two rows show the results for $\beta_{0}$ and $\beta_{1}$, and those
for $\beta_{2}$ are omitted to avoid redundancy. k-NNs and MGLS perform
better than FGLS, and this is in contrast to the performance of DGP1.
In general, MGLS works better than k-NNs except for a few cases.

To see the potential applicability of MGLS to a non-single index structure,
we consider another heteroskedastic linear model denoted by DGP4:
\begin{align}
Y_{i} & =\beta_{0}+\beta_{1}X_{1i}+\beta_{2}X_{2i}+u_{i},\quad u_{i}=\sigma_{i}\varepsilon_{i},\quad\varepsilon_{i}\sim N(0,1),\nonumber \\
\beta_{0} & =\beta_{1}=\beta_{2}=1,\quad\log(X_{1i}),\ \log(X_{2i})\sim N(0,1),\qquad X_{1i},\ X_{2i}\mbox{ and }\varepsilon_{i}\mbox{ are independent},\nonumber \\
\sigma_{i}^{2} & =.1+.2\tilde{X}_{i}+.3\tilde{X}_{i}^{2},\quad\log(\tilde{X}_{i})=\frac{\log(X_{1i})+\log(X_{2i})}{\sqrt{2}}.\label{eq:DGP4}
\end{align}
The right panel of Table \ref{tab:multivariate} shows the results.
The results for DGP 4 are overall similar to those of DGP3. An exception
is FGLS, which performs poorly for DGP3. MGLS works remarkably well
for the heteroskedasticity of a non-single index structure.

\begin{landscape} 
\begin{table}[htp]
{\small{}\caption{{\small{}\label{tab:multivariate}}Simulation: Estimation with multivariate
covariates}
}{\small\par}
\raggedright{}%
\begin{tabular}{|l|llcccc|cccccc|}
\hline 
 & \multicolumn{6}{c|}{DGP 3} & \multicolumn{6}{c|}{DGP 4}\tabularnewline
\hline 
 & \multicolumn{2}{c}{$n=50$} & \multicolumn{2}{c}{$n=100$} & \multicolumn{2}{c|}{$n=500$} & \multicolumn{2}{c}{$n=50$} & \multicolumn{2}{c}{$n=100$} & \multicolumn{2}{c|}{$n=500$}\tabularnewline
Estimator  & RMSE & MAE & RMSE & MAE & RMSE & MAE & RMSE & MAE & RMSE & MAE & RMSE & MAE\tabularnewline
\hline 
GLS (infeasible) & 0.162 & 0.103 & 0.110  & 0.071  & 0.045  & 0.028  & 0.165 & 0.108 & 0.115  & 0.076  & 0.049  & 0.033 \tabularnewline
 & 0.163 & 0.107 & 0.109  & 0.072  & 0.048  & 0.033  & 0.108 & 0.067 & 0.071  & 0.046  & 0.029  & 0.020 \tabularnewline
OLS  & 3.401 & 3.589 & 4.255  & 4.168  & 6.650  & 6.695  & 3.069 & 2.653 & 3.792  & 2.980  & 4.897  & 4.186 \tabularnewline
 & 1.942 & 1.950 & 2.317  & 2.260  & 3.051  & 2.610  & 2.318 & 2.170 & 2.914  & 2.338  & 3.809  & 3.198 \tabularnewline
FGLS  & 2.531 & 2.239 & 2.516  & 2.141  & 2.731  & 2.037  & 1.381 & 1.189 & 1.427  & 1.233  & 1.699  & 1.219 \tabularnewline
 & 1.606 & 1.441 & 1.709  & 1.486  & 1.732  & 1.358  & 1.359 & 1.227 & 1.344  & 1.281  & 1.326  & 1.271 \tabularnewline
k-NN (Automatic) & 1.952 & 1.925 & 2.108  & 1.709  & 1.786  & 1.537  & 1.868 & 1.638 & 1.778  & 1.488  & 1.709  & 1.390 \tabularnewline
 & 1.546 & 1.429 & 1.680  & 1.489  & 1.516  & 1.318  & 1.766 & 1.771 & 1.865  & 1.763  & 2.009  & 1.626 \tabularnewline
k-NN ($k=6$)  & 1.827 & 1.766 & 1.787  & 1.587  & 1.670  & 1.666  & 1.719 & 1.541 & 1.674  & 1.521  & 1.594  & 1.537 \tabularnewline
 & 1.458 & 1.397 & 1.514  & 1.486  & 1.497  & 1.362  & 1.717 & 1.704 & 1.769  & 1.764  & 1.813  & 1.654 \tabularnewline
k-NN ($k=15$)  & 1.914 & 1.957 & 1.850  & 1.669  & 1.490  & 1.385  & 1.769 & 1.611 & 1.669  & 1.491  & 1.373  & 1.246 \tabularnewline
 & 1.468 & 1.401 & 1.511  & 1.428  & 1.313  & 1.248  & 1.712 & 1.727 & 1.769  & 1.639  & 1.588  & 1.517 \tabularnewline
k-NN ($k=24$)  & 2.182 & 2.203 & 2.008  & 1.816  & 1.570  & 1.510  & 1.952 & 1.888 & 1.799  & 1.581  & 1.392  & 1.254 \tabularnewline
 & 1.562 & 1.455 & 1.611  & 1.571  & 1.371  & 1.267  & 1.825 & 1.807 & 1.890  & 1.729  & 1.626  & 1.511 \tabularnewline
MGLS  & 2.144 & 1.977 & 1.993  & 1.659  & 1.667  & 1.481  & 1.839 & 1.549 & 1.647  & 1.422  & 1.320  & 1.251 \tabularnewline
 & 1.486 & 1.467 & 1.477  & 1.401  & 1.238  & 1.186  & 1.670 & 1.533 & 1.604  & 1.451  & 1.448  & 1.360 \tabularnewline
\hline 
\end{tabular}\vspace{0.5cm}
\\
{\footnotesize{}Note: ``RMSE'' and ``MAE'' stand for the root mean
squared error and mean absolute error, respectively. The results for
GLS report}\\
{\footnotesize{}the levels of the RMSE and MAE, and those for others
are their ratios relative to GLS.}{\footnotesize\par}
\end{table}
 \end{landscape}

Next, we turn to the simulation results on inference. Tables \ref{tab:ci-univariate}
and \ref{tab:ci-multivariate} show empirical coverages (EC) and average
lengths (AL) for the 95\% confidence intervals under DGPs 1-4. Again
we consider GLS, OLS, FGLS, k-NN, and MGLS. For OLS, three types of
confidence intervals are considered. They are based on the usual OLS
standard error (OLS-U), the heteroskedasticity-robust standard error
(OLS-R), and the wild bootstrap standard error (OLS-boot). For MGLS,
we also present the results for its robust version. We observe that
the empirical coverages are smaller than the nominal coverage 0.95
for all DGPs and all methods except GLS. It is natural that OLS-U
performs poorly since it is invalid except for DGP2. The performance
of k-NN is worse than others for all DGPSs in terms of empirical coverage.
OLS-R, OLS-boot, FGLS, and MGLS work similarly in terms of empirical
coverage, however, we note that the average length of OLS-R is much
larger than those of FGLS and MGLS except for DGP2. While the empirical
coverages of OLS-Boot are similar to those of OLS-R, the average lengths
of OLS-Boot are smaller than those of OLS-R but still larger than
those of MGLS. MGLS works quite well for all DGPs, especially for
$n=500$. The results of MGLS (Robust) are similar to those of MGLS
especially when $n=100$ and 500. Finally, we note that the empirical
coverages tend to be lower when $n=50$ than when $n=100$ and $500$.
Careful interpretation of results is recommended when the sample size
is small.

\begin{landscape} 
\begin{table}[htp]
{\small{}\caption{{\small{}\label{tab:ci-univariate}}Simulation: Inference with univariate
covariate}
}{\small\par}

\begin{tabular}{|l|cccccc|cccccc|}
\hline 
 & \multicolumn{6}{c|}{DGP 1} & \multicolumn{6}{c|}{DGP 2}\tabularnewline
\hline 
 & \multicolumn{2}{c}{$n=50$} & \multicolumn{2}{c}{$n=100$} & \multicolumn{2}{c|}{$n=500$} & \multicolumn{2}{c}{$n=50$} & \multicolumn{2}{c}{$n=100$} & \multicolumn{2}{c|}{$n=500$}\tabularnewline
Estimator  & EC & AL & EC  & AL  & EC  & AL  & EC & AL & EC  & AL  & EC  & AL \tabularnewline
\hline 
GLS (infeasible) & 0.956 & 0.528 & 0.955  & 0.370  & 0.956  & 0.164  & 0.939 & 0.749 & 0.947  & 0.516  & 0.955  & 0.224 \tabularnewline
 & 0.946 & 0.627 & 0.962  & 0.441  & 0.960  & 0.196  & 0.948 & 0.316 & 0.952  & 0.207  & 0.970  & 0.085 \tabularnewline
OLS-U  & 0.798 & 1.008 & 0.742  & 0.740  & 0.636  & 0.349  & 0.939 & 0.749 & 0.947  & 0.516  & 0.955  & 0.224 \tabularnewline
 & 0.492 & 0.409 & 0.421  & 0.290  & 0.348  & 0.131  & 0.948 & 0.316 & 0.952  & 0.207  & 0.970  & 0.085 \tabularnewline
OLS-R  & 0.766 & 0.962 & 0.805  & 0.862  & 0.884  & 0.648  & 0.933 & 0.733 & 0.941  & 0.507  & 0.949  & 0.222 \tabularnewline
 & 0.730 & 0.761 & 0.772  & 0.689  & 0.880  & 0.488  & 0.874 & 0.272 & 0.881  & 0.185  & 0.935  & 0.081 \tabularnewline
OLS-Boot & 0.740 & 0.885 & 0.845 & 0.517 & 0.907 & 0.356 & 0.917 & 0.718 & 0.947 & 0.516 & 0.955 & 0.224\tabularnewline
 & 0.690 & 0.681 & 0.856 & 0.527 & 0.894 & 0.345 & 0.846 & 0.270 & 0.952 & 0.207 & 0.970 & 0.085\tabularnewline
FGLS  & 0.800 & 0.451 & 0.847  & 0.328  & 0.872  & 0.162  & 0.916 & 0.709 & 0.925  & 0.493  & 0.935  & 0.216 \tabularnewline
 & 0.737 & 0.504 & 0.812  & 0.395  & 0.885  & 0.195  & 0.761 & 0.231 & 0.758  & 0.159  & 0.844  & 0.072 \tabularnewline
k-NN (Automatic) & 0.708 & 0.410 & 0.659  & 0.258  & 0.701  & 0.102  & 0.902 & 0.711 & 0.884  & 0.483  & 0.883  & 0.205 \tabularnewline
 & 0.576 & 0.351 & 0.574  & 0.251  & 0.650  & 0.115  & 0.927 & 0.306 & 0.917  & 0.197  & 0.881  & 0.079 \tabularnewline
k-NN ($k=6$)  & 0.732 & 0.410 & 0.666  & 0.258  & 0.621  & 0.102  & 0.845 & 0.711 & 0.845  & 0.483  & 0.819  & 0.205 \tabularnewline
 & 0.576 & 0.351 & 0.574  & 0.251  & 0.650  & 0.115  & 0.927 & 0.306 & 0.917  & 0.197  & 0.881  & 0.079 \tabularnewline
k-NN ($k=15$)  & 0.735 & 0.418 & 0.704  & 0.266  & 0.717  & 0.105  & 0.929 & 0.725 & 0.907  & 0.492  & 0.914  & 0.210 \tabularnewline
 & 0.582 & 0.353 & 0.592  & 0.258  & 0.677  & 0.118  & 0.944 & 0.310 & 0.931  & 0.200  & 0.919  & 0.081 \tabularnewline
k-NN ($k=24$)  & 0.711 & 0.440 & 0.688  & 0.269  & 0.721  & 0.107  & 0.945 & 0.744 & 0.921  & 0.504  & 0.935  & 0.216 \tabularnewline
 & 0.512 & 0.324 & 0.537  & 0.244  & 0.668  & 0.118  & 0.953 & 0.316 & 0.942  & 0.204  & 0.939  & 0.083 \tabularnewline
MGLS  & 0.779 & 0.499 & 0.812  & 0.363  & 0.905  & 0.165  & 0.885 & 0.640 & 0.907  & 0.468  & 0.937  & 0.219 \tabularnewline
 & 0.725 & 0.523 & 0.744  & 0.392  & 0.888  & 0.188  & 0.951 & 0.333 & 0.968  & 0.222  & 0.972  & 0.092 \tabularnewline
MGLS (Robust) & 0.762 & 0.483 & 0.791 & 0.354 & 0.903 & 0.163 & 0.879 & 0.635 & 0.902 & 0.463 & 0.933 & 0.216\tabularnewline
 & 0.725 & 0.465 & 0.744 & 0.359 & 0.888 & 0.181 & 0.951 & 0.258 & 0.968 & 0.177 & 0.972 & 0.079\tabularnewline
\hline 
\end{tabular}
\raggedright{}\vspace{0.5cm}
\\
{\footnotesize{}Note: ``EC'' and ``AL'' stand for the empirical coverage
probability and average length, respectively. ``OLS-U'', ``OLS-R'',
and}\\
{\footnotesize{}``OLS-Boot'' use the normal approximation with the
usual OLS standard error, the heteroskedasticity robust standard error,}\\
{\footnotesize{}and the percentile bootstrap interval, respectively.
``MGLS (Robust)'' is based on the variance formula presented in Remark
2.}{\footnotesize\par}
\end{table}
 \end{landscape}

\begin{landscape} 
\begin{table}[htp]
{\small{}\caption{{\small{}\label{tab:ci-multivariate}}Simulation: Inference with multivariate
covariates}
}{\small\par}
\raggedright{}%
\begin{tabular}{|l|cccccc|cccccc|}
\hline 
 & \multicolumn{6}{c|}{DGP 3} & \multicolumn{6}{c|}{DGP 4}\tabularnewline
\hline 
 & \multicolumn{2}{c}{$n=50$} & \multicolumn{2}{c}{$n=100$} & \multicolumn{2}{c|}{$n=500$} & \multicolumn{2}{c}{$n=50$} & \multicolumn{2}{c}{$n=100$} & \multicolumn{2}{c|}{$n=500$}\tabularnewline
Estimator  & EC & AL & EC  & AL  & EC  & AL  & EC & AL & EC  & AL  & EC  & AL \tabularnewline
\hline 
GLS (infeasible)  & 0.944 & 0.611 & 0.951  & 0.413  & 0.946  & 0.175  & 0.944 & 0.636 & 0.943  & 0.440  & 0.956  & 0.192 \tabularnewline
 & 0.951 & 0.632 & 0.961  & 0.439  & 0.960  & 0.194  & 0.949 & 0.414 & 0.950  & 0.282  & 0.968  & 0.123 \tabularnewline
OLS-U  & 0.824 & 1.535 & 0.786  & 1.108  & 0.632  & 0.511  & 0.819 & 1.222 & 0.780  & 0.893  & 0.675  & 0.412 \tabularnewline
 & 0.589 & 0.526 & 0.549  & 0.369  & 0.491  & 0.164  & 0.639 & 0.420 & 0.611  & 0.298  & 0.521  & 0.133 \tabularnewline
OLS-R  & 0.787 & 1.411 & 0.815  & 1.232  & 0.869  & 0.873  & 0.797 & 1.197 & 0.843  & 1.068  & 0.906  & 0.718 \tabularnewline
 & 0.729 & 0.767 & 0.782  & 0.660  & 0.891  & 0.441  & 0.762 & 0.596 & 0.810  & 0.517  & 0.914  & 0.334 \tabularnewline
OLS-Boot & 0.756 & 1.319 & 0.781 & 1.133 & 0.839 & 0.797 & 0.752 & 1.115 & 0.785 & 0.951 & 0.860 & 0.654\tabularnewline
 & 0.688 & 0.708 & 0.749 & 0.593 & 0.826 & 0.400 & 0.719 & 0.569 & 0.780 & 0.460 & 0.866 & 0.301\tabularnewline
FGLS  & 0.831 & 1.069 & 0.845  & 0.759  & 0.897  & 0.336  & 0.801 & 0.596 & 0.823  & 0.424  & 0.834  & 0.191 \tabularnewline
 & 0.658 & 0.517 & 0.722  & 0.395  & 0.826  & 0.198  & 0.797 & 0.382 & 0.862  & 0.262  & 0.855  & 0.112 \tabularnewline
k-NN (Automatic) & 0.571 & 0.481 & 0.557  & 0.289  & 0.587  & 0.105  & 0.672 & 0.526 & 0.646  & 0.323  & 0.609  & 0.122 \tabularnewline
 & 0.471 & 0.296 & 0.472  & 0.205  & 0.534  & 0.091  & 0.596 & 0.298 & 0.599  & 0.200  & 0.605  & 0.082 \tabularnewline
k-NN ($k=6$)  & 0.597 & 0.481 & 0.574  & 0.289  & 0.549  & 0.105  & 0.670 & 0.526 & 0.639  & 0.323  & 0.571  & 0.122 \tabularnewline
 & 0.471 & 0.296 & 0.472  & 0.205  & 0.534  & 0.091  & 0.596 & 0.298 & 0.599  & 0.200  & 0.605  & 0.082 \tabularnewline
k-NN ($k=15$)  & 0.592 & 0.504 & 0.590  & 0.299  & 0.639  & 0.108  & 0.690 & 0.551 & 0.675  & 0.338  & 0.689  & 0.129 \tabularnewline
 & 0.491 & 0.301 & 0.484  & 0.212  & 0.570  & 0.096  & 0.607 & 0.309 & 0.629  & 0.210  & 0.668  & 0.088 \tabularnewline
k-NN ($k=24$)  & 0.582 & 0.557 & 0.561  & 0.313  & 0.618  & 0.111  & 0.681 & 0.590 & 0.664  & 0.347  & 0.702  & 0.132 \tabularnewline
 & 0.459 & 0.287 & 0.450  & 0.204  & 0.562  & 0.096  & 0.598 & 0.297 & 0.608  & 0.207  & 0.662  & 0.091 \tabularnewline
MGLS  & 0.803 & 0.956 & 0.863  & 0.623  & 0.938  & 0.248  & 0.801 & 0.805 & 0.844  & 0.526  & 0.902  & 0.216 \tabularnewline
 & 0.687 & 0.524 & 0.756  & 0.401  & 0.897  & 0.198  & 0.707 & 0.404 & 0.776  & 0.305  & 0.908  & 0.154 \tabularnewline
MGLS (Robust) & 0.755 & 0.855 & 0.833 & 0.573 & 0.920 & 0.234 & 0.762 & 0.750 & 0.833 & 0.511 & 0.902 & 0.219\tabularnewline
 & 0.687 & 0.548 & 0.756 & 0.415 & 0.897 & 0.199 & 0.707 & 0.422 & 0.776 & 0.308 & 0.908 & 0.148\tabularnewline
\hline 
\end{tabular}\vspace{0.5cm}
\\
{\footnotesize{}Note: ``EC'' and ``AL'' stand for the empirical coverage
probability and average length, respectively. ``OLS-U'', ``OLS-R'',
and}\\
{\footnotesize{}``OLS-Boot'' use the normal approximation with the
usual OLS standard error, the heteroskedasticity robust standard error,}\\
{\footnotesize{}and the percentile bootstrap interval, respectively.
``MGLS (Robust)'' is based on the variance formula presented in Remark
2.}{\footnotesize\par}
\end{table}
 \end{landscape}

\subsection{Empirical example\label{sub:emp}}

We illustrate how the proposed method in this paper can improve the
precision of the traditional OLS approach. In doing so, we revisit
Acemoglu and Restrepo (2017) that investigate the relationship between
an aging population and economic growth. After Hansen (1939), a popular
perspective is that countries undergoing faster aging suffer more
economically partly because of excessive savings by an aging population.
In contrast to the perspective, Acemoglu and Restrepo (2017) find
no evidence of a negative relationship between aging and GDP per capita
after controlling for initial GDP per capita, initial demographic
composition, and trends by region.

Acemoglu and Restrepo (2017) estimated eight specifications for the
regression of the change in (log) GDP per capita from 1990 to 2015
(denoted by GDP) on the population aging measured by the change in
the ratio of the population above 50 to those between the ages of
20 and 49 (denoted by Aging). The results are reproduced in Panel
A of Table \ref{tab:empirical}. Those in columns 1-5 are based on
the sample including 169 countries. Column 1 shows the result of the
simple regression. Standard errors robust to heteroskedasticity are
reported in square brackets. Column 2 shows the result with an additional
regressor, the initial log GDP per worker in 1990. Column 3 in addition
includes the initial demographic information, the ratio of the population
above 50 to those between 20 and 49 in 1990 (denoted by Initial Ratio),
and the population in 1990. Column 4 additionally uses dummies for
seven regions, Latin America, East Asia, South Asia, Africa, North
Africa and Middle East, Eastern Europe and Central Asia, and Developed
Countries. Column 5 estimates the same specification as Column 4 with
instruments of birthrates for the 1960, 1965, 1970, 1975, and 1980
cohorts. Columns 6 to 8 report the result for OECD countries using
specifications of Columns 1, 3, and 5, respectively. The number of
observations for the first five columns is 169, and that for the last
three columns is 35. Seven out of eight OLS estimates indicate positive
relationships and five of them are statistically significant at the
5 percent level. Acemoglu and Restrepo (2017) discuss that these findings
can be explained by the adoption of automation technologies based
on a theoretical model.

\begin{figure}[htbp]
\begin{centering}
\includegraphics[scale=0.45]{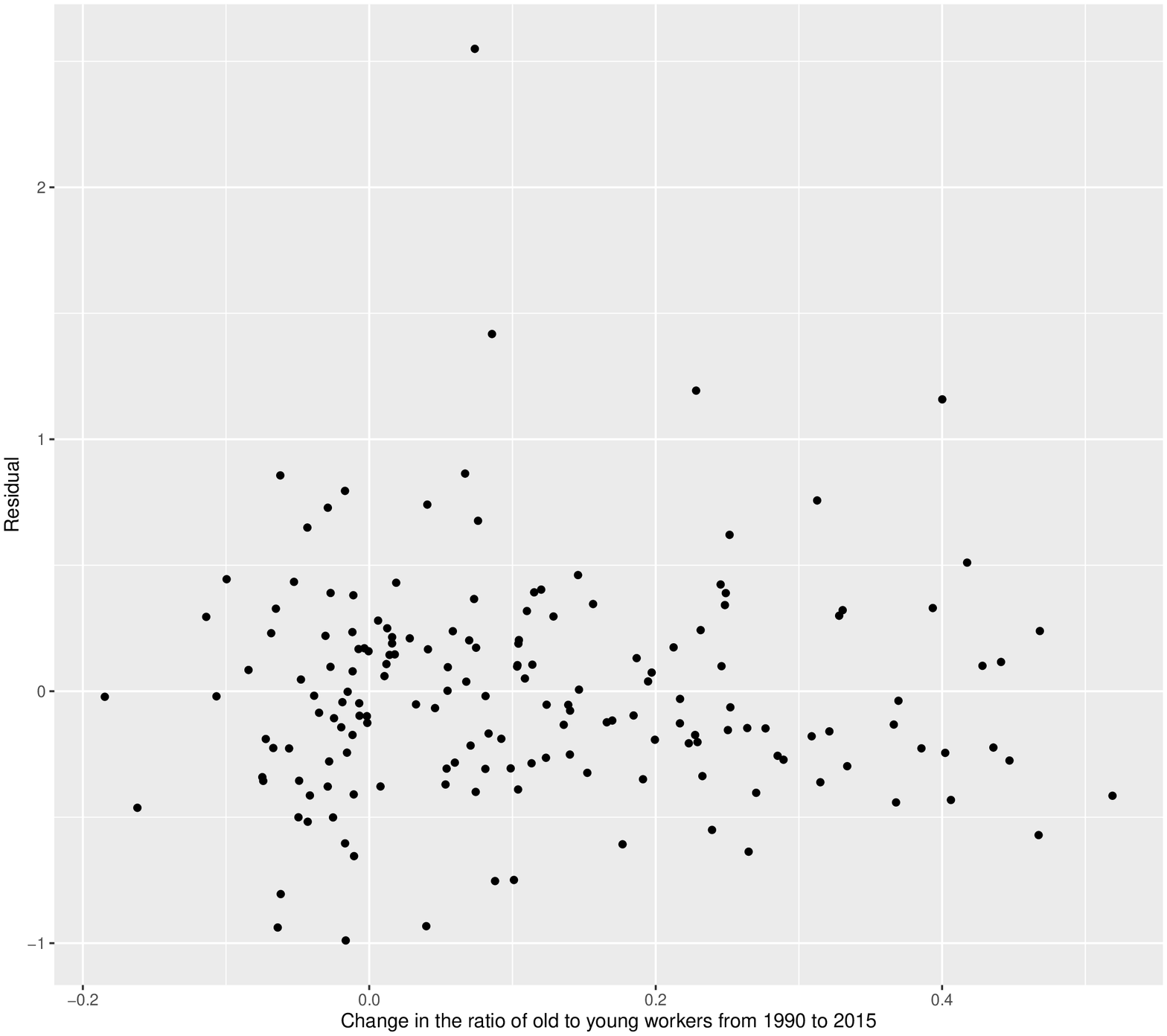} \includegraphics[scale=0.45]{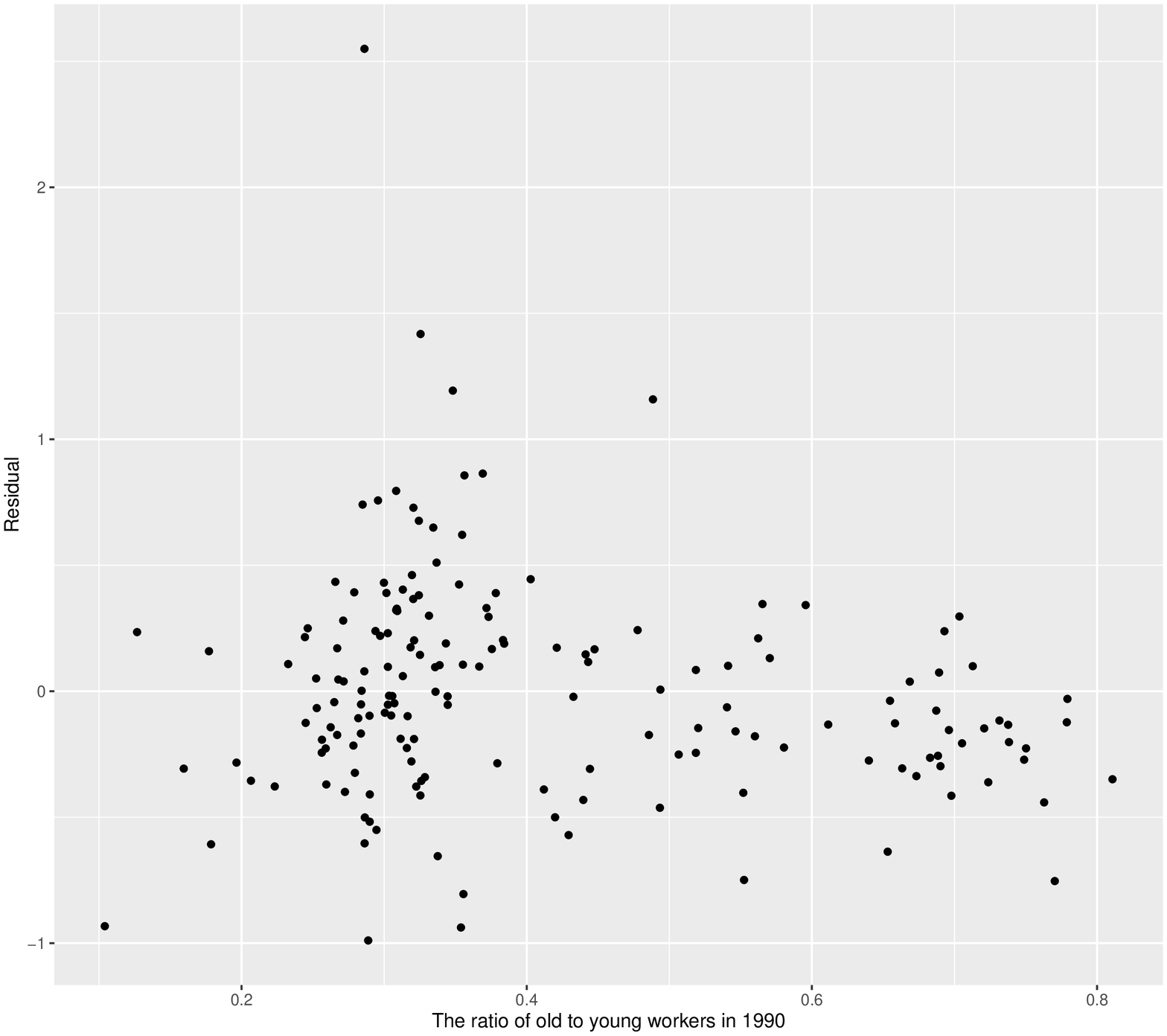}
\par\end{centering}
\begin{centering}
\caption{Plots for residual and aging (upper) and residual and ratio of old
to young workers in 1990 (lower)}
\par\end{centering}
$\quad$

$\quad$
\raggedright{}{\footnotesize{}Note: For both panels, the residuals
are obtained from the regression of the change in GDP per capita from
1990 to 2015 (GDP) on the population aging measured by the ratio of
the population above 50 to those between the ages of 20 and 49 (Aging).
For the upper panel, the variable on the X-axis represents the change
in the ratio of old to young workers from 1990 to 2015. For the lower
panel, it represents the ratio of old to young workers in 1990. }\label{fig:change} 
\end{figure}

We estimate the same specifications by MGLS proposed in this paper.
Acemoglu and Restrepo (2017) show that the negative effect of aging
can be mitigated or reversed by adopting new automation technologies
given abundant capital. This also implies that the effect of aging
can be negative without sufficient capital. Hence it would be reasonable
to consider Aging as a source of heteroskedasticity. The upper panel
of Figure \ref{fig:change} shows the relationship between the residual
from the simple regression of column 1 in Panel A and Aging. Heteroskedasticity
due to Aging is not easily confirmed visually. We consider Initial
Ratio as another source of heteroskedasticity since the low ratio
of old to young in 1990 is likely correlated with more aging in 2015,
leading to larger variability in GDP per capita by the same reasoning
discussed above. The lower panel of Figure \ref{fig:change} presents
the relationship between the residual from the simple regression of
column 1 in Panel A and Initial Ratio, and we see that the variability
decreases with the growing ratio. 

Panels B, C, and D of Table \ref{tab:empirical} show the results
of MGLS. Panels B and C present the results for cases where the conditional
error variance functions depend on Aging and Initial Ratio, respectively.
Panel D reports the results where the conditional error variance functions
depend on all exogenous regressors except the regional dummies. Standard
errors based on Theorems 1-2 and their analogous versions for IV estimators
are reported in parentheses, while robust standard errors are reported
in square brackets. First, we observe reductions in standard errors
for all MGLSs relative to OLS. The differences stand out when $n=169$.
Second, the two standard errors are similar for the MGLS estimates
under exogeneity while they differ for the IV estimates. These are
the supporting evidence for the monotone specification of the conditional
error variance function for the MGLS method with exogenous regressors
but not for the IV method. Third, the results given in Columns 2,
3, and 4 are stable, while the results of IV estimates and OECD countries
contain a lot of variations. Those variations can be due to non-monotone
conditional error variance functions and/or small sample sizes, and
further investigations will be required. Overall, the standard errors
of MGLS tend to be smaller or no larger than those of OLS, which demonstrates
the increased precision of MGLS. 

\begin{landscape} 
\begin{table}[htp]
\caption{Effects of the Aging on GDP by OLS and MGLS}

\begin{centering}
\begin{tabular}{|c|rrrrc|rrc|}
\hline 
 & \multicolumn{5}{c|}{Sample of all countries ($n=169$)} & \multicolumn{3}{c|}{OECD countries ($n=35$)}\tabularnewline
\hline 
Specification  & \multicolumn{1}{c}{(1)} & \multicolumn{1}{c}{(2)} & \multicolumn{1}{c}{(3)} & \multicolumn{1}{c}{(4)} & (5) & \multicolumn{1}{c}{(6)} & \multicolumn{1}{c}{(7)} & (8)\tabularnewline
\hline 
Panel A: OLS  &  &  &  &  &  &  &  & \tabularnewline
Aging & 0.335 & 1.036 & 1.162 & 0.773 & 1.703 & -0.262 & 0.042 & 1.186\tabularnewline
 & (0.210) & (0.257) & (0.276) & (0.322) & (0.411) & (0.352) & (0.346) & (0.458)\tabularnewline
Initial GDP  &  & -0.153 & -0.138 & -0.156 & -0.190 &  & -0.205 & -0.260\tabularnewline
 &  & (0.039) & (0.042) & (0.046) & (0.045) &  & (0.072) & (0.092)\tabularnewline
\hline 
Panel B: MGLS (covariate of $\sigma^{2}(\cdot)=$Aging)  &  &  &  &  &  &  &  & \tabularnewline
Aging & 0.387 & 1.098 & 1.191 & 0.751 & 0.414 & -0.391 & -0.029 & -0.458\tabularnewline
 & (0.189) & (0.187) & (0.205) & (0.267) & (0.101) & (0.247) & (0.284) & (0.092)\tabularnewline
 & {[}0.150{]} & {[}0.179{]} & {[}0.198{]} & {[}0.310{]} & {[}0.472{]} & {[}0.190{]} & {[}0.340{]} & {[}0.451{]}\tabularnewline
Initial GDP &  & -0.164 & -0.155 & -0.168 & -0.079 &  & -0.190 & -0.297\tabularnewline
 &  & (0.027) & (0.029) & (0.030) & (0.009) &  & (0.069) & (0.025)\tabularnewline
 &  & {[}0.031{]} & {[}0.032{]} & {[}0.029{]} & {[}0.046{]} &  & {[}0.069{]} & {[}0.141{]}\tabularnewline
\hline 
Panel C: MGLS (covariate of $\sigma^{2}(\cdot)=$Initial Ratio)  &  &  &  &  &  &  &  & \tabularnewline
Aging & 0.065 & 0.771 & 0.894 & 0.574 & 0.483 & -0.501 & -0.344 & -0.585\tabularnewline
 & (0.196) & (0.223) & (0.231) & (0.235) & (0.142) & (0.270) & (0.219) & (0.226)\tabularnewline
 & {[}0.196{]} & {[}0.249{]} & {[}0.262{]} & {[}0.272{]} & {[}0.603{]} & {[}0.231{]} & {[}0.213{]} & {[}0.758{]}\tabularnewline
Initial GDP &  & -0.164 & -0.141 & -0.159 & -0.080 &  & -0.148 & -0.379\tabularnewline
 &  & (0.031) & (0.035) & (0.041) & (0.012) &  & (0.056) & (0.096)\tabularnewline
 &  & {[}0.035{]} & {[}0.037{]} & {[}0.046{]} & {[}0.055{]} &  & {[}0.065{]} & {[}0.288{]}\tabularnewline
\hline 
Panel D: MGLS (covariates of $\sigma^{2}(\cdot)=$All) &  &  &  &  &  &  &  & \tabularnewline
Aging & 0.285 & 1.064 & 1.188 & 0.810 & 0.494 & -0.391 & 0.062 & -0.268\tabularnewline
 & (0.221) & (0.265) & (0.281) & (0.289) & (0.100) & (0.247) & (0.274) & (0.186)\tabularnewline
 & {[}0.206{]} & {[}0.249{]} & {[}0.271{]} & {[}0.323{]} & {[}0.442{]} & {[}0.190{]} & {[}0.340{]} & {[}0.972{]}\tabularnewline
Initial GDP &  & -0.152 & -0.136 & -0.146 & -0.079 &  & -0.203 & -0.250\tabularnewline
 &  & (0.030) & (0.033) & (0.041) & (0.009) &  & (0.072) & (0.049)\tabularnewline
 &  & {[}0.033{]} & {[}0.036{]} & {[}0.044{]} & {[}0.051{]} &  & {[}0.072{]} & {[}0.197{]}\tabularnewline
\hline 
\end{tabular}
\par\end{centering}
\label{tab:empirical}\vspace{0.5cm}

\raggedright{}{\footnotesize{}Note: For all specifications from (1)
to (8), GDP is the dependent variable. Column 1 shows the result of
the simple regression of Aging on GDP. Column 2 shows the result with
an additional regressor, the initial log GDP per worker in 1990. Column
3, in addition, includes the initial demographic information, the
ratio of the population above 50 to those between 20 and 49 in 1990
(denoted by Initial Ratio), and the population in 1990. Column 4 additionally
uses dummies for seven regions, Latin America, East Asia, South Asia,
Africa, North Africa and Middle East, Eastern Europe and Central Asia,
and Developed Countries. Columns (6), (7) and (8) report the result
for OECD countries using specifications (1), (3) and (5), respectively.
Panel A reproduces the results by Acemoglu and Restrepo (2017). For
Panel A, heteroskedasticity robust standard errors are presented in
parentheses. Panels B, C, and D present the results by MGLS. Panels
B and C show the results where the conditional error variance functions
depend on Aging and Initial Ratio, respectively. Panel D reports the
results where the conditional error variance functions depend on all
exogenous variables except the regional dummies. For Columns (1)-(4)
and (6)-(7) of Panels B and C, standard errors based on the formula
in Theorem 1 are presented in parentheses, while those based on the
formula in Remark 3 are presented in square brackets. For Columns
(1)-(4) and (6)-(7) of Panel D, standard errors based on the formula
in Theorem 2 are presented in parentheses, while those based on the
formula analogous to Remark 3 are presented in square brackets. For
Columns (5) and (8) of Panels B, C, and D, standard errors are based
on the formulae analogous to Remark 3.}{\footnotesize\par}
\end{table}
 \end{landscape}

\newpage{}

\appendix

\section{Proof of lemma and theorem in Section \ref{sec:case1}\label{app:uni-proofs}}

\textbf{Notation. }In this section, we use the following notation.
For a function $f(\cdot)$, we let $\left\Vert f\right\Vert _{\infty}=\sup_{x\in\mathscr{X}}|f(x)|$
be the sup-norm and $\left\Vert f\right\Vert _{2,P}=\sqrt{\int|f(x)|^{2}dP}$
be the $L_{2}(P)$ norm; given there is no confusion in the context,
we use the same set of notations for a vector $a=(a_{1},\dots,a_{k})^{\prime}$:
we let $\left\Vert a\right\Vert _{\infty}=\max_{j\in\{1,\ldots,k\}}|a_{k}|$
be the sup-norm and $\left\Vert a\right\Vert =\sqrt{\sum_{j=1}^{k}a_{j}^{2}}$
be the Euclidean norm. $D_{A}^{L}[f](a)$ be the left derivative of
the greatest convex minorant of a function $f$ evaluated at $a\in A$,
$\mathbb{P}_{n}$ be the empirical measure of $\{Y_{i},X_{i},Z_{i}\}_{i=1}^{n}$,
$\mathbb{G}_{n}$ be the empirical process, i.e., $\mathbb{G}_{n}f=\frac{1}{\sqrt{n}}\sum_{i=1}^{n}\{f(X_{i})-E[f(X_{i})]\}$,
$\left\Vert \mathbb{G}_{n}\right\Vert _{\mathcal{F}}=\sup_{f\in\mathcal{F}}|\mathbb{G}_{n}f|$,
and $\mathbb{I}_{A}(x)=\mathbb{I}\{x\in A\}$. Let $\tau_{0}(x)=\sigma^{2}(x)$,
$\tau_{0}^{\prime}(x_{L})$ be the right derivative of $\tau_{0}$
at $x_{L}$, $\hat{\tau}(x)=\hat{\sigma}^{2}(x)$, $\mathscr{W}$
be the support of $W:=(1,X,Z^{\prime})^{\prime}$, $F(x)$ be the
distribution function of $X$, $F_{n}(x)=\frac{1}{n}\sum_{i=1}^{n}\mathbb{I}\{X_{i}\le x\}$,
and $M_{n}(x)=\frac{1}{n}\sum_{i=1}^{n}\hat{U}_{i}^{2}\mathbb{I}\{X_{i}\le x\}$.
For $a,b\in\mathbb{R}$, let $a\wedge b$ denote $\min\{a,b\}$, and
$a\lesssim b$ denote that there exists a positive constant $C$ such
that $a\leq C\cdot b$. Let $\text{dim}(w)$ be the dimension of a
vector $w$. 

\subsection{Proof of Lemma \ref{lem:trun_uni}\label{app-sub:lem1}}

Since $\hat{U}_{j}=Y_{j}-W_{j}^{\prime}\hat{\theta}_{\mathrm{OLS}}$
is the OLS residual, Assumptions A1-A2 and $\hat{\theta}_{\mathrm{OLS}}-\theta=O_{p}(n^{-1/2})$
imply $\hat{U}_{j}^{2}-U_{j}^{2}=O_{p}(n^{-1/2}\log n)=o_{p}(n^{-1/3})$
uniformly over $j=1,\ldots,n$. To see this, decompose
\begin{eqnarray*}
\hat{U}_{j}^{2}-U_{j}^{2} & = & (Y_{j}-W_{j}^{\prime}\hat{\theta}_{\mathrm{OLS}})^{2}-(Y_{j}-W_{j}^{\prime}\theta)^{2}\\
 & = & W_{j}^{\prime}(\hat{\theta}_{\mathrm{OLS}}+\theta)\cdot W_{j}^{\prime}(\hat{\theta}_{\mathrm{OLS}}-\theta)-2W_{j}^{\prime}\theta\cdot W_{j}^{\prime}(\hat{\theta}_{\mathrm{OLS}}-\theta)-2U_{j}W_{j}^{\prime}(\hat{\theta}_{\mathrm{OLS}}-\theta)\\
 & =: & I_{j}+II_{j}+III_{j}.
\end{eqnarray*}
For $I_{j}$, note that
\begin{eqnarray}
I_{j} & = & [W_{j}^{\prime}(\hat{\theta}_{\mathrm{OLS}}-\theta)]^{2}+2W_{j}^{\prime}\theta\cdot W_{j}^{\prime}(\hat{\theta}_{\mathrm{OLS}}-\theta)\nonumber \\
 & \leq & \left\Vert W_{j}\right\Vert ^{2}\left\Vert \hat{\theta}_{\mathrm{OLS}}-\theta\right\Vert ^{2}+2\left\Vert W_{j}\right\Vert \cdot\left\Vert \theta\right\Vert \cdot\left\Vert W_{j}\right\Vert \cdot\left\Vert \hat{\theta}_{\mathrm{OLS}}-\theta\right\Vert \nonumber \\
 & \le & R^{2}\left\Vert \hat{\theta}_{\mathrm{OLS}}-\theta\right\Vert ^{2}+2R^{2}\left\Vert \theta\right\Vert \cdot\left\Vert \hat{\theta}_{\mathrm{OLS}}-\theta\right\Vert =O_{p}(n^{-1/2}),\label{eq:I_j}
\end{eqnarray}
where $R$ is the constant defined in Assumption A1. The first inequality
follows from the Cauchy-Schwarz inequality, the second inequality
follows from $\left\Vert W_{j}\right\Vert \le R$ (by Assumption A1),
and the last equality follows from $\hat{\theta}_{\mathrm{OLS}}-\theta=O_{p}(n^{-1/2})$.
Note that in the second inequality, the upper bound no longer depends
on the index $j$, so we have $\max_{j}|I_{j}|=O_{p}(n^{-1/2})$.
For $II_{j}$, using the same reasoning as for the first inequality
in (\ref{eq:I_j}), we have $\max_{j}|II_{j}|=O_{p}(n^{-1/2})$. Note
that here we only consider the second term following the first inequality
of (\ref{eq:I_j}). For $III_{j}$, the same argument as above yields
$\max_{j}|W_{j}^{\prime}(\hat{\theta}_{\mathrm{OLS}}-\theta)|=O_{p}(n^{-1/2})$.
Furthermore, by Assumption A2 and a similar argument after equation
(7.11) on p.3297 of Balabdaoui, Durot and Jankowski (2019) (BDJ hereafter),
we have $\max_{1\le j\le n}|U_{j}^{2}|=O_{p}(\log n)$. By the fact
that $\max_{1\le j\le n}|U_{j}|\leq\max_{1\le j\le n}|U_{j}^{2}$|
if $\max_{1\le j\le n}|U_{j}|\geq1$, we have 
\begin{equation}
\max_{1\le j\le n}|U_{j}|=O_{p}(\log n).\label{eq:max_Uj}
\end{equation}
In the case of $\max_{1\le j\le n}|U_{j}|<1$, $\max_{1\le j\le n}|U_{j}|=O_{p}(\log n)$
holds trivially. Combining (\ref{eq:max_Uj}) and $\max_{j}|W_{j}^{\prime}(\hat{\theta}_{\mathrm{OLS}}-\theta)|=O_{p}(n^{-1/2})$,
we have $\max_{j}|III_{j}|=O_{p}(n^{-1/2}\log n)$. Consequently,
we have
\begin{eqnarray}
\max_{j}|\hat{U}_{j}^{2}-U_{j}^{2}| & \leq & \max_{j}|I_{j}|+\max_{j}|II_{j}|+\max_{j}|III_{j}|\nonumber \\
 & = & O_{p}(n^{-1/2}\log n)=o_{p}(n^{-1/3}).\label{eq:U_op1/3}
\end{eqnarray}

Furthermore, Assumption A3 guarantees $q_{n}^{*}-x_{L}=O(n^{-1/3})$
(by an expansion of $q_{n}^{*}=F^{-1}(n^{-1/3})$ for the quantile
function $F^{-1}(\cdot)$ of $X$), and we can define $c^{*}=\lim_{n\to\infty}n^{1/3}(q_{n}^{*}-x_{L})=\left.\frac{dF^{-1}(q)}{dq}\right|_{q\downarrow0}\in(0,\infty)$.

Now, we analyze $n^{1/3}\{\hat{\tau}(q_{n}^{*})-\tau_{0}(x_{L})\}$.
The term $n^{1/3}\{\hat{\tau}(q_{n})-\tau_{0}(q_{n})\}$ will be addressed
in the final step of this subsection. Pick any $m>0$. Let 
\[
Z_{n1}(t)=n^{2/3}[\{n^{-1/3}m+\tau_{0}(x_{L})\}F_{n}(x_{L}+t(q_{n}^{*}-x_{L}))-M_{n}(x_{L}+t(q_{n}^{*}-x_{L}))].
\]
Observe that 
\begin{eqnarray}
 &  & P\left(n^{1/3}\{\hat{\tau}(q_{n}^{*})-\tau_{0}(x_{L})\}\leq m\right)=P\left(\arg\underset{s\in[x_{L},x_{U}]}{\max}[\{n^{-1/3}m+\tau_{0}(x_{L})\}F_{n}(s)-M_{n}(s)]\geq q_{n}^{*}\right)\nonumber \\
 & = & P\left(\arg\underset{t\in[0,(x_{U}-x_{L})/(q_{n}^{*}-x_{L})]}{\max}n^{-2/3}Z_{n1}(t)\geq1\right),\label{pf:obj_switch}
\end{eqnarray}
where the first equality follows from the switch relation (see a review
by Groeneboom and Jongbloed, 2014), and the second equality follows
from a change of variables $s=x_{L}+t(q_{n}^{*}-x_{L})$ and its implication,
$s\geq q_{n}^{*}\Leftrightarrow t\geq1$. Let $\hat{U}(y,w)=y-w^{\prime}\hat{\theta}_{\mathrm{OLS}}$
and 
\[
g_{n,t}(y,w)=n^{1/6}\{\tau_{0}(x_{L})-\hat{U}(y,w)^{2}\}\mathbb{I}_{[x_{L},x_{L}+t(q_{n}^{*}-x_{L})]}(x).
\]
We decompose
\begin{eqnarray*}
Z_{n1}(t) & = & \sqrt{n}(\mathbb{P}_{n}-P)g_{n,t}+n^{2/3}E[\{\tau_{0}(x_{L})-\hat{U}(Y,W)^{2}\}\mathbb{I}_{[x_{L},x_{L}+t(q_{n}^{*}-x_{L})]}(X)]\\
 &  & +n^{1/3}m\{F_{n}(x_{L}+t(q_{n}^{*}-x_{L}))-F(x_{L}+t(q_{n}^{*}-x_{L}))\}+n^{1/3}mF(x_{L}+t(q_{n}^{*}-x_{L}))\\
 & =: & Z_{n1}^{a}(t)+Z_{n1}^{b}(t)+Z_{n1}^{c}(t)+Z_{n1}^{d}(t).
\end{eqnarray*}

\textbf{Analysis of $Z_{n1}^{a}(t)$.} We verify the conditions of
van der Vaart (2000, Theorem 19.28). Define the class of random functions
(depending on $\hat{\theta}_{\mathrm{OLS}}$): 
\[
\mathcal{G}_{n1}=\{g_{n,t}(y,w)=n^{1/6}(\tau_{0}(x_{L})-\hat{U}(y,w)^{2})\mathbb{I}_{[x_{L},x_{L}+t(q_{n}^{*}-x_{L})]}(x):t\in[0,K]\},
\]
for $K\in(0,\infty)$, where $n$ in the subscript indicates the dependence
on both the scaling parameter $n^{1/6}$ and $\hat{\theta}_{\mathrm{OLS}}$.
By van der Vaart (2000, Example 19.6) we know that for a bracket size
$\epsilon$, $\mathcal{G}_{n1}$ has the entropy with bracketing of
order $\log(1/\epsilon)$. Thus, $\mathcal{G}_{n1}$ satisfies the
entropy condition for van der Vaart (2000, Theorem 19.28).

For each $t,s\in[0,K]$, note that 
\begin{eqnarray}
\mathrm{Cov}(g_{n,t},g_{n,s}) & = & n^{1/3}E[\{\hat{U}(Y,W)^{2}-\tau_{0}(x_{L})\}^{2}\mathbb{I}_{[x_{L},x_{L}+(t\land s)(q_{n}^{*}-x_{L})]}(X)]+o_{p}(1)\nonumber \\
 & = & n^{1/3}E[\{U^{2}-\tau_{0}(x_{L})\}^{2}\mathbb{I}_{[x_{L},x_{L}+(t\land s)(q_{n}^{*}-x_{L})]}(X)]+o_{p}(1)\nonumber \\
 & = & n^{1/3}E[[\varepsilon^{2}+\{\tau_{0}(X)-\tau_{0}(x_{L})\}^{2}]\mathbb{I}_{[x_{L},x_{L}+(t\land s)(q_{n}^{*}-x_{L})]}(X)]+o_{p}(1)\nonumber \\
 & = & n^{1/3}\int_{x_{L}}^{x_{L}+(t\land s)(q_{n}^{*}-x_{L})}[\sigma_{\varepsilon}^{2}(x)+\{\tau_{0}(x)-\tau_{0}(x_{L})\}^{2}]f_{X}(x)dx+o_{p}(1)\nonumber \\
 & = & [\sigma_{\varepsilon}^{2}(\xi_{n})+\{\tau_{0}(\xi_{n})-\tau_{0}(x_{L})\}^{2}]f_{X}(\xi_{n})c^{*}(t\land s)+o_{p}(1)\nonumber \\
 & = & \sigma_{\varepsilon}^{2}(x_{L})f_{X}(x_{L})c^{*}(t\land s)+o_{p}(1),\label{pf:cov}
\end{eqnarray}
for $\xi_{n}\in(x_{L},x_{L}+(t\land s)q_{n}^{*})$. The first equality
follows from $q_{n}^{*}-x_{L}=O(n^{-1/3})$. In the second equality,
we replace the estimated $\hat{U}^{2}$ with the unobservable $U^{2}$.
By (\ref{eq:U_op1/3}), the discrepancy between $\hat{U}^{2}$ and
$U^{2}$ converges more rapidly than $n^{-1/3}$, and the factor $\mathbb{I}_{[x_{L},x_{L}+(t\land s)(q_{n}^{*}-x_{L})]}(X)$
further refines this rate. Consequently, under Assumptions A1 and
A2, the impact of substituting $\hat{U}^{2}$ with $U^{2}$ in the
second line is $o_{p}(1)$. The third equality follows from the definition
$\varepsilon=U^{2}-\tau_{0}(X)$ and $E[\varepsilon|X]=0$, the fourth
equality follows from the law of iterated expectations, the fifth
equality follows from a Taylor expansion, and the last equality follows
from $c^{*}=\lim_{n\to\infty}n^{1/3}(q_{n}^{*}-x_{L})$ and the continuity
of $\sigma_{\varepsilon}^{2}(\cdot)$ and $\tau_{0}(\cdot)$ at $x_{L}$
from right. Similarly, we have $\mathrm{Var}(g_{n,t})=\sigma_{\varepsilon}^{2}(x_{L})f_{X}(x_{L})c^{*}t+o_{p}(1)$.

We next consider the envelop function of the class $\mathcal{G}_{n1}$,
that is 
\[
G_{n1}(y,w)=n^{1/6}|\tau_{0}(x_{L})-\hat{U}(y,w)^{2}|\cdot\mathbb{I}_{[x_{L},x_{L}+K(q_{n}^{*}-x_{L})]}(x).
\]
We can see that $G_{n1}$ is square integrable since similar arguments
to (\ref{pf:cov}) yield
\begin{eqnarray}
E[G_{n1}^{2}(Y,W)] & = & n^{1/3}E[|\tau_{0}(x_{L})-\hat{U}(Y,W)^{2}|\cdot\mathbb{I}_{[x_{L},x_{L}+K(q_{n}^{*}-x_{L})]}(X)]\nonumber \\
 & = & n^{1/3}E[|\tau_{0}(x_{L})-U^{2}|\cdot\mathbb{I}_{[x_{L},x_{L}+K(q_{n}^{*}-x_{L})]}(X)]+o_{p}(1)\nonumber \\
 & = & n^{1/3}E[[\varepsilon^{2}+\{\tau_{0}(X)-\tau_{0}(x_{L})\}^{2}]\cdot\mathbb{I}_{[x_{L},x_{L}+K(q_{n}^{*}-x_{L})]}(X)]+o_{p}(1)\nonumber \\
 & = & n^{1/3}\int_{x_{L}}^{x_{L}+K(q_{n}^{*}-x_{L})}[\sigma_{\varepsilon}^{2}(x)+\{\tau_{0}(x)-\tau_{0}(x_{L})\}^{2}]f_{X}(x)dx+o_{p}(1)\nonumber \\
 & = & O_{p}(1),\label{pf:envelop_si}
\end{eqnarray}
and thus the Lindeberg condition can be verified by Assumption A2:
for any $\zeta>0$ and some $\delta>0$, 
\begin{eqnarray}
E[G_{n1}^{2}\mathbb{I}\{G_{n1}>\zeta\sqrt{n}\}] & \leq & \frac{n^{(2+\delta)1/6}}{\zeta^{\delta}n^{\delta/2}}E[|\tau_{0}(x_{L})-\hat{U}(Y,W)^{2}|^{2+\delta}\cdot\mathbb{I}_{[x_{L},x_{L}+K(q_{n}^{*}-x_{L})]}(X)]\nonumber \\
 & = & \frac{n^{(2+\delta)1/6}}{\zeta^{\delta}n^{\delta/2}}E[|\tau_{0}(x_{L})-U^{2}|^{2+\delta}\cdot\mathbb{I}_{[x_{L},x_{L}+K(q_{n}^{*}-x_{L})]}(X)]+o_{p}(1)\nonumber \\
 & = & O(n^{-\delta/3})+o_{p}(1)=o_{p}(1),\label{pf:envelop_Linde}
\end{eqnarray}
where the inequality follows from the same arguments that are used
in the proof of Markov's inequality, the first equality follows from
$\hat{\theta}_{\mathrm{OLS}}-\theta=O_{p}(n^{-1/2})$ and Assumptions
A1-A2, and the second equality follows from a similar argument to
(\ref{pf:envelop_si}). 

Furthermore, as $\delta_{n}\to0$, we obtain 
\begin{eqnarray}
\underset{|t-s|\leq\delta_{n}}{\sup}E|g_{n,t}-g_{n,s}|^{2} & = & n^{1/3}\underset{|t-s|\leq\delta_{n}}{\sup}E[\{\hat{U}(Y,W)^{2}-\tau_{0}(x_{L})\}^{2}\mathbb{I}_{[x_{L},x_{L}+|t-s|q_{n}^{*}]}(X)]\nonumber \\
 & = & n^{1/3}\underset{|t-s|\leq\delta_{n}}{\sup}E[[\varepsilon^{2}+\{\tau_{0}(X)-\tau_{0}(x_{L})\}^{2}]\cdot\mathbb{I}_{[x_{L},x_{L}+|t-s|q_{n}^{*}]}(X)]+o_{p}(\delta_{n})\nonumber \\
 & = & O_{p}(\delta_{n})=o_{p}(1).\label{pf:equicon}
\end{eqnarray}

By (\ref{pf:cov})-(\ref{pf:equicon}), we can apply van der Vaart
(2000, Theorem 19.28), which implies for each $K\in(0,\infty)$, 
\begin{equation}
Z_{n1}^{a}(t)\overset{d}{\to}\sqrt{\sigma_{\varepsilon}^{2}(x_{L})f_{X}(x_{L})c^{*}}\mathcal{W}_{t}\text{ in }l^{\infty}[0,K].\label{pf:Za}
\end{equation}

\textbf{Analysis of $Z_{n1}^{b}(t)$.} Observe that 
\begin{eqnarray}
Z_{n1}^{b}(t) & = & n^{2/3}E[\{\tau_{0}(x_{L})-U^{2}\}\mathbb{I}{}_{[x_{L},x_{L}+t(q_{n}^{*}-x_{L})]}(X)]+n^{2/3}E[(U^{2}-\hat{U}(Y,W)^{2})\mathbb{I}{}_{[x_{L},x_{L}+t(q_{n}^{*}-x_{L})]}(X)]\nonumber \\
 & = & n^{2/3}\int_{x_{L}}^{x_{L}+t(q_{n}^{*}-x_{L})}\{\tau_{0}(x_{L})-\tau_{0}(F^{-1}(F(x)))\}dF(x)+o_{p}(1)\nonumber \\
 & = & n^{2/3}\int_{F(x_{L})}^{F(x_{L}+t(q_{n}^{*}-x_{L}))}\{\tau_{0}(x_{L})-\tau_{0}(F^{-1}(v))\}dv+o_{p}(1)\nonumber \\
 & = & -n^{2/3}\int_{F(x_{L})}^{F(x_{L}+t(q_{n}^{*}-x_{L}))}\tau_{0}^{\prime}(x_{L})\{F^{-1}(v)-F^{-1}(F(x_{L}))\}dv+o_{p}(1)\nonumber \\
 & = & -n^{2/3}\int_{F(x_{L})}^{F(x_{L}+t(q_{n}^{*}-x_{L}))}\tau_{0}^{\prime}(x_{L})\frac{v-F(x_{L})}{f_{X}(x_{L})}dv+o_{p}(1)\nonumber \\
 & = & -n^{2/3}\tau_{0}^{\prime}(x_{L})\frac{\{F(x_{L}+t(q_{n}^{*}-x_{L}))-F(x_{L})\}^{2}}{2f_{X}(x_{L})}+o_{p}(1)\nonumber \\
\text{} & = & -\tau_{0}^{\prime}(x_{L})\frac{t^{2}(c^{*})^{2}}{2}f_{X}(x_{L})+o_{p}(1)\label{pf:Zb}
\end{eqnarray}
holds uniformly over $t\in[0,K]$, where the second equality follows
from $E[\{U^{2}-\hat{U}(Y,W)^{2}\}\cdot\mathbb{I}{}_{[x_{L},x_{L}+t(q_{n}^{*}-x_{L})]}(X)]=o_{p}(n^{-2/3})$,
the third equality follows from a change of variables $v=F(x)$, the
fourth equality follows from a Taylor expansion, the fifth equality
follows from $F^{-1}(v)-x_{L}=\frac{1}{f_{X}(x_{L})}(v-F(x_{L}))+o(v-F(x_{L}))$,
the sixth equality follows from evaluating the integral, and the last
equality follows from a Taylor expansion and $c^{*}=\lim_{n\to\infty}n^{1/3}(q_{n}^{*}-x_{L})$.

\textbf{Analysis of $Z_{n1}^{c}(t)$.} By Kim and Pollard (1990, Maximal
inequality 3.1), 
\[
E\left[\underset{t\in[0,K]}{\sup}|F_{n}(x_{L}+t(q_{n}^{*}-x_{L}))-F(x_{L}+t(q_{n}^{*}-x_{L}))|\right]\leq n^{-1/2}J\sqrt{PG_{n}^{2}}
\]
holds for some constant $J\in(0,\infty)$. Here $G_{n}$ is the envelope
of the set of indicator functions, thus $PG_{n}^{2}\leq1$. As a
result,
\begin{equation}
Z_{n1}^{c}(t)\le n^{1/3}n^{-1/2}mJ\sqrt{PG_{n}^{2}}=o(1),\label{pf:Zc}
\end{equation}
uniformly over $t\in[0,K]$.

\textbf{Analysis of $Z_{n1}^{d}(t)$.} A Taylor expansion yields 
\begin{equation}
Z_{n1}^{d}(t)=n^{1/3}mF(x_{L}+t(q_{n}^{*}-x_{L}))=m\cdot t\cdot f_{X}(x_{L})c^{*}+o(1),\label{pf:Zd}
\end{equation}
uniformly over $t\in[0,K]$, for every $K<\infty$.

Combining (\ref{pf:Za})-(\ref{pf:Zd}), it holds that for each $0<K<\infty$,
\begin{equation}
Z_{n1}(t)\overset{d}{\to}Z_{1}(t):=\sqrt{\sigma_{\varepsilon}^{2}(x_{L})f_{X}(x_{L})c^{*}}\mathcal{W}_{t}-\tau_{0}^{\prime}(x_{L})\frac{t^{2}(c^{*})^{2}}{2}f_{X}(x_{L})+m\cdot t\cdot f_{X}(x_{L})c^{*}\text{ in }l^{\infty}[0,K].\label{eq: Limit_Z}
\end{equation}

We now verify the conditions of the argmax continuous mapping theorem
(Kim and Pollard, 1990). Note that for each $t\ne s$, 
\[
\mathrm{Var}(Z_{1}(s)-Z_{1}(t))=\sigma_{\varepsilon}^{2}(x_{L})f_{X}(x_{L})c^{*}|t-s|\ne0.
\]
By Kim and Pollard (1990), the process $t\to Z_{1}(t)$ achieves its
maximum a.s. at a unique point. Consider extended versions of $Z_{n1}$
and $Z_{1}$ to the real line: 
\[
\tilde{Z}_{n1}(t)=\begin{cases}
Z_{n1}(t), & t\geq0\\
t & t<0
\end{cases},\qquad\tilde{Z}_{1}(t)=\begin{cases}
Z_{1}(t), & t\geq0\\
t & t<0
\end{cases}.
\]
It holds $\tilde{Z}_{n1}(t)\overset{d}{\to}\tilde{Z}_{1}(t)$, and
the similar argument to Lemma SM.2.1 (ii) in Babii and Kumar (2023)
yields that the maximum of $\tilde{Z}_{n1}(t)$ is uniformly tight.
Therefore, by Kim and Pollard (1990, Theorem 2.7), 
\begin{eqnarray*}
 &  & P\left(n^{1/3}\{\hat{\tau}(q_{n}^{*})-\tau_{0}(x_{L})\}\leq m\right)\to P\left(\left[\arg\underset{t\ge0}{\max}Z_{1}(t)\right]\geq1\right)\\
 & = & P\left(\left[\arg\underset{t\ge0}{\max}\sqrt{\frac{\sigma_{\varepsilon}^{2}(x_{L})}{c^{*}f_{X}(x_{L})}}\mathcal{W}_{t}-\tau_{0}^{\prime}(x_{L})\frac{t^{2}c^{*}}{2}+mt\right]\geq1\right)\\
 & = & P\left(\left[D_{[0,\infty)}^{L}\left(\sqrt{\frac{\sigma_{\varepsilon}^{2}(x_{L})}{c^{*}f_{X}(x_{L})}}\mathcal{W}_{t}+\tau_{0}^{\prime}(x_{L})\frac{t^{2}c^{*}}{2}\right)(1)\right]\leq m\right),
\end{eqnarray*}
where the second equality follows from the switch relation and symmetry
of the process $\mathcal{W}_{t}$. Thus, we have 
\begin{equation}
n^{1/3}\{\hat{\tau}(q_{n}^{*})-\tau_{0}(x_{L})\}\overset{d}{\to}D_{[0,\infty)}^{L}\left(\sqrt{\frac{\sigma_{\varepsilon}^{2}(x_{L})}{c^{*}f_{X}(x_{L})}}\mathcal{W}_{t}+\tau_{0}^{\prime}(x_{L})\frac{t^{2}c^{*}}{2}\right)(1),\label{pf:rate_qxl}
\end{equation}
which also implies 
\begin{eqnarray}
 &  & n^{1/3}\{\hat{\tau}(q_{n}^{*})-\tau_{0}(q_{n}^{*})\}\nonumber \\
 & \overset{d}{\to} & D_{[0,\infty)}^{L}\left(\sqrt{\frac{\sigma_{\varepsilon}^{2}(x_{L})}{c^{*}f_{X}(x_{L})}}\mathcal{W}_{t}+\tau_{0}^{\prime}(x_{L})\frac{t^{2}c^{*}}{2}\right)(1)-\underset{n\to\infty}{\lim}n^{1/3}\{\tau_{0}(q_{n}^{*})-\tau_{0}(x_{L})\}\nonumber \\
 & \overset{d}{\sim} & D_{[0,\infty)}^{L}\left(\sqrt{\frac{\sigma_{\varepsilon}^{2}(x_{L})}{c^{*}f_{X}(x_{L})}}\mathcal{W}_{t}+\tau_{0}^{\prime}(x_{L})\frac{t^{2}c^{*}}{2}-\tau_{0}^{\prime}(x_{L})c^{*}t\right)(1),\label{pf:concl}
\end{eqnarray}
where the distribution relation follows from the fact that the $D_{[0,\infty)}^{L}$
is a linear operator for a linear function of $t$.

Finally, we analyze $n^{1/3}\{\hat{\tau}(q_{n})-\tau_{0}(q_{n})\}$.
Recall $q_{n}$ is the $(n^{-1/3})$-th sample quantile of $X$. Assumption
A3 guarantees $q_{n}-q_{n}^{*}=O_{p}(n^{-1/2})=o_{p}(n^{-1/3})$,
which also implies $\text{plim}_{n\to\infty}n^{1/3}(q_{n}-x_{L})=\lim_{n\to\infty}n^{1/3}(q_{n}^{*}-x_{L})=c^{*}$.
Thus, the same argument for (\ref{pf:rate_qxl}) can be applied to
show that the result in (\ref{pf:rate_qxl}) holds true even if we
replace $q_{n}^{*}$ with $q_{n}$. Therefore, the conclusion follows.

\subsection{Proof of Theorem \ref{thm:uni}\label{app-sub:thm_uni}}

By the definitions of the estimators, it holds that
\begin{eqnarray*}
\sqrt{n}(\hat{\theta}-\theta) & = & \left(\frac{1}{n}\sum_{i:x_{i}>q_{n}}\hat{\sigma}_{i}^{-2}W_{i}W_{i}^{\prime}\right)^{-1}\left(\frac{1}{\sqrt{n}}\sum_{i:x_{i}>q_{n}}\hat{\sigma}_{i}^{-2}W_{i}U_{i}\right),\\
\sqrt{n}(\hat{\theta}_{\mathrm{IGLS}}-\theta) & = & \left(\frac{1}{n}\sum_{i=1}^{n}\sigma_{i}^{-2}W_{i}W_{i}^{\prime}\right)^{-1}\left(\frac{1}{\sqrt{n}}\sum_{i=1}^{n}\sigma_{i}^{-2}W_{i}U_{i}\right).
\end{eqnarray*}
Thus it is sufficient for the conclusion to show 
\begin{eqnarray*}
T_{1} & := & \frac{1}{\sqrt{n}}\sum_{i:x_{i}>q_{n}}\hat{\sigma}_{i}^{-2}W_{i}U_{i}-\frac{1}{\sqrt{n}}\sum_{i=1}^{n}\sigma_{i}^{-2}W_{i}U_{i}\overset{p}{\to}0,\\
T_{2} & := & \frac{1}{n}\sum_{i:x_{i}>q_{n}}\hat{\sigma}_{i}^{-2}W_{i}W_{i}^{\prime}-\frac{1}{n}\sum_{i=1}^{n}\sigma_{i}^{-2}W_{i}W_{i}^{\prime}\overset{p}{\to}0.
\end{eqnarray*}

\subsubsection{The concentration of $T_{1}$}

Decompose 
\[
T_{1}=\frac{1}{\sqrt{n}}\sum_{i:x_{i}>q_{n}}(\hat{\sigma}_{i}^{-2}-\sigma_{i}^{-2})W_{i}U_{i}-\frac{1}{\sqrt{n}}\sum_{i:x_{i}\leq q_{n}}\sigma_{i}^{-2}W_{i}U_{i}=:T_{11}-T_{12}.
\]

We first consider $T_{12}$. For any $h\in\{1:\text{dim}(W)\}$, let
$W_{i}^{h}$ and $T_{12}^{h}$ be the $h$-th element of $W_{i}$
and $T_{12}$, respectively. Note that $E[T_{12}^{h}|q_{n}]=0$ by
$E[U|W]=0$. Also we have $\mathrm{Var}(T_{12}^{h}|q_{n})\overset{p}{\to}0$.
To see this, decompose
\[
\mathrm{Var}(T_{12}^{h}|q_{n})=I_{h}-n\cdot(II_{h})^{2},
\]
where $I_{h}=\frac{1}{n}E\left[\left.\left(\sum_{i=1}^{n}\mathbb{I}\{X_{i}\leq q_{n}\}\sigma_{i}^{-2}W_{i}^{h}U_{i}\right)^{2}\right|q_{n}\right]$
and $II_{h}=E[\mathbb{I}\{X_{i}\leq q_{n}\}\sigma_{i}^{-2}W_{i}^{h}U_{i}|q_{n}]$.
For $I_{h}$, note that
\begin{eqnarray*}
I_{h} & = & \frac{1}{n}E\left[\left.E\left[\left.\left(\sum_{i=1}^{n}\mathbb{I}\{X_{i}\leq q_{n}\}\sigma_{i}^{-2}W_{i}^{h}U_{i}\right)^{2}\right|\mathbf{W}\right]\right|q_{n}\right]\\
 & = & E[E[(\mathbb{I}\{X_{i}\leq q_{n}\}\sigma_{i}^{-2}W_{i}^{h}U_{i})^{2}|\mathbf{W}]|q_{n}]=E[\mathbb{I}\{X\leq q_{n}\}\sigma^{-2}(X)(W^{h})^{2}|q_{n}]\\
 & \leq & R^{2}\sigma^{-2}(x_{L})E[\mathbb{I}\{X\leq q_{n}\}|q_{n}]\overset{p}{\to}0,
\end{eqnarray*}
where $\mathbf{W}=(W_{1},\dots,W_{n})^{\prime}$. The first equality
follows from the law of iterated expectation and the fact that $q_{n}$
is a function of $\mathbf{W}$, the second equality follows from $E[U|W]=0$
and $\{U_{i}\}_{i=1}^{n}$ being iid, the third equality follows because
conditional on $\mathbf{W}$, $\mathbb{I}\{X_{i}\leq q_{n}\}(\sigma_{i}^{-2}W_{i}^{h})^{2}$
is treated as fixed, the inequality follows from Assumptions A1 and
A2, and the convergence follows from $q_{n}\overset{p}{\to}x_{L}$.
For $II_{h}$, note that
\[
II_{h}=E[\mathbb{I}\{X_{i}\leq q_{n}\}\sigma_{i}^{-2}W_{i}^{h}E[U_{i}|\mathbf{W}]|q_{n}]=0,
\]
where the first equality follows from the law of iterated expectation
and the fact that $q_{n}$ is a function of $\mathbf{W}$, and the
second equality follows from $E[U_{i}|\mathbf{W}]=E[U_{i}|W_{i}]=0$.
Since $E[T_{12}^{h}|q_{n}]=0$ and $\mathrm{Var}(T_{12}^{h}|q_{n})\overset{p}{\to}0$
hold for every $h$, we can conclude that $T_{12}\overset{p}{\to}0$.

To proceed, we will utilize Lemma \ref{lem: t1_uni} below. Its proof
can be found at the end of Appendix \ref{app-sub:thm_uni}. Recall
that earlier in this appendix, we relabel $\sigma^{2}(\cdot)$ as
$\tau_{0}(\cdot)$, and $\hat{\tau}$ is used to denote the isotonic
estimator of $\sigma^{2}(\cdot)$. Additionally, with some abuse of
notation, we use $w_{h}$ to denote the $h$-th element of vector
$w$.

\begin{lem} \label{lem: t1_uni} Under Assumptions A1-A3, 
\begin{description}
\item [{(i)}] $\left\Vert \hat{\tau}\right\Vert _{\infty}=O_{p}(\log n)$, 
\item [{(ii)}] $\left\Vert \hat{\tau}-\tau_{0}\right\Vert _{2,P}^{2}=O_{p}((\log n)^{2}n^{-2/3})$, 
\item [{(iii)}] $E[\left\Vert \mathbb{G}_{n}\right\Vert _{\mathcal{F}_{n}}]\le\frac{A\nu}{2}$
holds for any constants $A>0$ and $\nu>0$, and all sufficiently
large $n$, where $\mathcal{F}_{n}$ is the function class defined
as
\begin{equation}
\mathcal{F}_{n}=\left\{ f_{n}(w,u)=\mathbb{I}\{x>q_{n}\}\left(\frac{1}{\tau(x)}-\frac{1}{\tau_{0}(x)}\right)w_{h}u:\begin{array}{c}
\tau\ge0\text{ is monotone increasing on }\mathscr{X},\\
\left\Vert \tau\right\Vert _{\infty}\le C\log n,\text{ }\left\Vert \tau-\tau_{0}\right\Vert _{2,P}^{2}\le Cr_{n},\\
\mathbb{I}\{x>q_{n}\}/\tau(x)\leq1/K_{0},h\in\{1:\mathrm{dim}(w)\}
\end{array}\text{ }\right\} ,\label{eq: class_T1}
\end{equation}
with $C$ and $K_{0}$ being some positive constants, and $r_{n}=(\log n)^{2}n^{-2/3}$.
\end{description}
\end{lem}

Now we focus on $T_{11}$. Since the proof is similar, we only present
the proof for the $h$-th element of $T_{11}$, i.e., for any constant
$A>0$, 
\begin{equation}
P\{|\mathbb{G}_{n}\hat{f}|\ge A\}\to0,\label{pf:A1}
\end{equation}
where $\hat{f}(w,u)=\mathbb{I}\{x>q_{n}\}\left(\frac{1}{\hat{\tau}(x)}-\frac{1}{\tau_{0}(x)}\right)w_{h}u$.
To this end, we set $\tau_{0}(x_{L})=C_{0}=2K_{0}>0$. It holds that
for any $A>0$ and $\nu>0$, there exists a positive constant $C$
such that 
\begin{eqnarray}
P\{|\mathbb{G}_{n}\hat{f}|\ge A\} & \le & P\left\{ |\mathbb{G}_{n}\hat{f}|\ge A,\text{ }\left\Vert \hat{\tau}\right\Vert _{\infty}\le C\log n,\text{ }\left\Vert \hat{\tau}-\tau_{0}\right\Vert _{2,P}^{2}\le Cr_{n},\text{ }\frac{\mathbb{I}\{x>q_{n}\}}{\hat{\tau}(x)}\leq\frac{1}{K_{0}}\right\} +\frac{\nu}{2}\nonumber \\
 & \le & \frac{E\left[\left\Vert \mathbb{G}_{n}\right\Vert _{\mathcal{F}_{n}}\right]}{A}+\frac{\nu}{2}\le\nu,\label{pf:main_inequality}
\end{eqnarray}
for all sufficiently large $n$, where the first inequality follows
from Lemma \ref{lem:trun_uni} and Lemma \ref{lem: t1_uni} (i)-(ii),
and the fact that $\hat{\tau}$ is monotone increasing (so that the
lower bound at the truncation point is the uniform lower bound). Specifically,
for any $\nu>0$, we can find $C>0$ and a positive integer $n_{0}$
such that for any integer $n>n_{0}$, it holds that (a) $P\{\left\Vert \hat{\tau}\right\Vert _{\infty}>C\log n\}<\frac{\nu}{6}$,
(b) $P\{\left\Vert \hat{\tau}-\tau_{0}\right\Vert _{2,P}^{2}>Cr_{n}\}<\frac{\nu}{6}$,
and (c) $P\left\{ \frac{\mathbb{I}\{x>q_{n}\}}{\hat{\tau}(x)}>\frac{1}{K_{0}}\right\} <\frac{\nu}{6}$.
Parts (a) and (b) are ensured by Lemma \ref{lem: t1_uni} (i) and
(ii), respectively; part (c) is guaranteed by Lemma \ref{lem:trun_uni}.
As a result, $P\left(\{\left\Vert \hat{\tau}\right\Vert _{\infty}>C\log n\}\text{ or }\{\left\Vert \hat{\tau}-\tau_{0}\right\Vert _{2,P}^{2}>Cr_{n}\}\text{ or }\left\{ \frac{\mathbb{I}\{x>q_{n}\}}{\hat{\tau}(x)}>\frac{1}{K_{0}}\right\} \right)<\frac{\nu}{2}$.
In the case of $\lim_{x\downarrow x_{L}}\frac{d\sigma^{2}(x)}{dx}=0$,
part (c) remains valid since $\hat{\tau}(x)$ will converge to $\tau_{0}$
at a faster rate (the $\sqrt{n}$-rate), then the first inequality
of (\ref{pf:main_inequality}) holds without invoking Lemma \ref{lem:trun_uni}.

The second inequality of (\ref{pf:main_inequality}) follows from
Markov's inequality and the definition of $\mathcal{F}_{n}$, which
is given by (\ref{eq: class_T1}). The last inequality follows from
Lemma \ref{lem: t1_uni} (iii). Since $\nu$ can be arbitrarily small,
we obtain (\ref{pf:A1}) and the conclusion follows.

\subsubsection{Proof of $T_{2}=o_{p}(1)$}

Note that
\begin{eqnarray*}
T_{2} & = & \frac{1}{n}\sum_{i:x_{i}>q_{n}}\hat{\sigma}_{i}^{-2}W_{i}W_{i}^{\prime}-\frac{1}{n}\sum_{i=1}^{n}\sigma_{i}^{-2}W_{i}W_{i}^{\prime}\\
 & = & \frac{1}{n}\sum_{i:x_{i}>q_{n}}(\hat{\sigma}_{i}^{-2}-\sigma_{i}^{-2})W_{i}W_{i}^{\prime}-\frac{1}{n}\sum_{i:x_{i}\leq q_{n}}\sigma_{i}^{-2}W_{i}W_{i}^{\prime}\\
 & =: & T_{21}-T_{22}.
\end{eqnarray*}
First, we have $T_{22}\overset{p}{\to}0$ since $q_{n}\overset{p}{\to}x_{L}$.
For $T_{21}$, let $s_{n}$ be the $(1-n^{-1/3})$-th sample quantile
of $\{X_{i}\}_{i=1}^{n}$. By employing arguments similar to those
in the proof of Lemma \ref{lem:trun_uni}, we have $\hat{\sigma}^{2}(s_{n})-\sigma^{2}(s_{n})=O_{p}(n^{-1/3})$.
Using reasoning akin to, yet simpler than, those in the proof of Lemma
\ref{lem:trun_uni}, we can establish that for any $x\in(q_{n},s_{n})$,
it holds that $\hat{\sigma}^{2}(x)-\sigma^{2}(x)=O_{p}(n^{-1/3})$.
Combining the aforementioned results with the monotonicity of both
$\hat{\sigma}^{2}(\cdot)$ and $\sigma^{2}(\cdot)$, we can conclude
that $\sup_{x\in[q_{n},s_{n}]}|\hat{\sigma}^{2}(x)-\sigma^{2}(x)|=O_{p}(n^{-1/3})$,
i.e., $\hat{\sigma}^{2}(x)$ is uniformly consistent within trimmed
domain $[q_{n},s_{n}]$ (the proof here resembles the one given for
the Glivenko-Cantelli Theorem regarding the uniform consistency of
the empirical distribution function; see, for example, the proof of
Theorem 19.1 in van der Vaart, 2000). Therefore, we have
\begin{equation}
T_{21}=\frac{1}{n}\sum_{i:q_{n}<x_{i}<s_{n}}(\hat{\sigma}_{i}^{-2}-\sigma_{i}^{-2})W_{i}W_{i}^{\prime}+\frac{1}{n}\sum_{i:x_{i}\geq s_{n}}(\hat{\sigma}_{i}^{-2}-\sigma_{i}^{-2})W_{i}W_{i}^{\prime}=o_{p}(1),\label{eq:conv_T21}
\end{equation}
where the second equality follows from the preceding argument, $|s_{n}-x_{U}|=O_{p}(n^{-1/3})$,
and Lemma \ref{lem: t1_uni} (i). Combining $T_{22}\overset{p}{\to}0$
and (\ref{eq:conv_T21}), we have $T_{2}\overset{p}{\to}0$.

\subsubsection{Proof of Lemma \ref{lem: t1_uni} (i)}

The min-max formula of the isotonic regression says 
\[
\min_{1\le k\le n}\frac{\sum_{j=1}^{k}\hat{U}_{j}^{2}}{k}\le\hat{\tau}(x)\le\max_{1\le k\le n}\frac{\sum_{j=k}^{n}\hat{U}_{j}^{2}}{n-k+1},
\]
for each $x\in\mathscr{X}$, which implies $\min_{1\le j\le n}\hat{U}_{j}^{2}\le\hat{\tau}(x)\le\max_{1\le j\le n}\hat{U}_{j}^{2}$
for each $x\in\mathscr{X}$. Thus, it is sufficient for the conclusion
to show that 
\begin{equation}
\max_{1\le j\le n}\hat{U}_{j}^{2}=O_{p}(\log n).\label{pf:1-1}
\end{equation}
Observe that 
\[
\max_{1\le j\le n}\hat{U}_{j}^{2}\le\max_{1\le j\le n}U_{j}^{2}+2Rk||\hat{\theta}_{\mathrm{OLS}}-\theta||_{\infty}\max_{1\le j\le n}|U_{j}|+R^{2}k^{2}||\hat{\theta}_{\mathrm{OLS}}-\theta||_{\infty}^{2}.
\]
From BDJ (2019, eq. (7.11) on p.3297), Assumption A2 guarantees $\max_{1\le j\le n}U_{j}^{2}=O_{p}(\log n)$.
Since $\hat{\theta}_{\mathrm{OLS}}$ is the OLS estimator, it holds
that $||\hat{\theta}_{\mathrm{OLS}}-\theta||_{\infty}=O_{p}(n^{-1/2})$.
By \eqref{eq:max_Uj}, we also have $\max_{1\le j\le n}|U_{j}|=O_{p}(\log n)$.
Combining these results with Assumption A1, we have (\ref{pf:1-1}).

\subsubsection{Proof of Lemma \ref{lem: t1_uni} (ii)}

The proof is based on that of Proposition 4 of BGH (p.8 of BGH-supp).
Recall that $\hat{\tau}(\cdot)$ is the solution of $\min_{\tau\in\{\text{all monotone funcitons}\}}\sum_{j=1}^{n}\{\hat{U}_{j}^{2}-\tau(X_{j})\}^{2}$,
or equivalently 
\begin{equation}
\max_{\tau\in\{\text{all monotone funcitons}\}}\sum_{j=1}^{n}\{2\hat{U}_{j}^{2}\tau(X_{j})-\tau(X_{j})\}.\label{pf:2-1}
\end{equation}
On the other hand, $\tau_{0}(\cdot)$ is the solution of $\min_{\tau\in\{\text{all monotone funcitons}\}}E[\{U^{2}-\tau(X)\}^{2}]$,
or equivalently 
\begin{equation}
\max_{\tau\in\{\text{all monotone funcitons}\}}E[2U^{2}\tau(X)-\tau(X)^{2}].\label{pf:2-2}
\end{equation}
By (\ref{pf:2-1}), it holds 
\[
\sum_{j=1}^{n}\{2\hat{U}_{j}^{2}\hat{\tau}(X_{j})-\hat{\tau}(X_{j})^{2}\}\ge\sum_{j=1}^{n}\{2\hat{U}_{j}^{2}\tau_{0}(X_{j})-\tau_{0}(X_{j})^{2}\},
\]
or equivalently (by plugging in $\hat{U}_{j}=U_{j}-W_{j}^{\prime}(\hat{\theta}_{\mathrm{OLS}}-\theta)$),
\begin{eqnarray}
 &  & \sum_{j=1}^{n}\{2U_{j}^{2}\hat{\tau}(X_{j})-\hat{\tau}(X_{j})^{2}\}+2\sum_{j=1}^{n}\left(-2U_{j}W_{j}^{\prime}(\hat{\theta}_{\mathrm{OLS}}-\theta)+\{W_{j}^{\prime}(\hat{\theta}_{\mathrm{OLS}}-\theta)\}^{2}\right)\{\hat{\tau}(X_{j})-\tau_{0}(X_{j})\}\nonumber \\
 & \ge & \sum_{j=1}^{n}\{2U_{j}^{2}\tau_{0}(X_{j})-\tau_{0}(X_{j})^{2}\}.\label{pf:2-3}
\end{eqnarray}
Define $d_{2}^{2}(\tau_{1},\tau_{2})=-E[2\tau_{1}\tau_{2}-\tau_{1}^{2}-\tau_{2}^{2}]$.
Note that for any monotone function $\tau$, 
\begin{eqnarray}
 &  & E[2U^{2}\tau(X)-\tau(X)^{2}]-E[2U^{2}\tau_{0}(X)-\tau_{0}(X)^{2}]\nonumber \\
 & = & E[2E[U^{2}|X]\tau(X)-\tau(X)^{2}-2E[U^{2}|X]\tau_{0}(X)+\tau_{0}(X)^{2}]\nonumber \\
 & = & E[2\tau_{0}(X)\tau(X)-\tau(X)^{2}-\tau_{0}(X)^{2}]=-d_{2}^{2}(\tau,\tau_{0}),\label{pf:2-4}
\end{eqnarray}
where the first equality follows from the law of iterated expectation,
the second equality follows from the definition $\tau_{0}(x)=E[U^{2}|X=x]$,
and the last equality follows from the definition of $d_{2}^{2}(\cdot,\cdot)$.

Define 
\begin{eqnarray*}
g_{\tau}(u,x) & = & \{2u^{2}\tau(x)-\tau(x)^{2}\}-\{2u^{2}\tau_{0}(x)-\tau_{0}(x)^{2}\},\\
R_{n} & = & \frac{2}{n}\sum_{j=1}^{n}\left(-2U_{j}W_{j}^{\prime}(\hat{\theta}_{\mathrm{OLS}}-\theta)+\{W_{j}^{\prime}(\hat{\theta}_{\mathrm{OLS}}-\theta)\}^{2}\right)\{\hat{\tau}(X_{j})-\tau_{0}(X_{j})\}.
\end{eqnarray*}
From (\ref{pf:2-3}) and (\ref{pf:2-4}), it holds 
\begin{equation}
\int g_{\hat{\tau}}(u,x)d(\mathbb{P}_{n}-P)(u,x)+R_{n}\ge d_{2}^{2}(\hat{\tau},\tau_{0}).\label{pf:2-5}
\end{equation}
Note that $R_{n}$ is bounded as
\[
|R_{n}|\le\left|-(\hat{\theta}_{\mathrm{OLS}}-\theta)^{\prime}\frac{4}{n}\sum_{j=1}^{n}W_{j}U_{j}\{\hat{\tau}(X_{j})-\tau_{0}(X_{j})\}\right|+\left|\frac{2}{n}\sum_{j=1}^{n}\{W_{j}^{\prime}(\hat{\theta}_{\mathrm{OLS}}-\theta)\}^{2}\{\hat{\tau}(X_{j})-\tau_{0}(X_{j})\}\right|.
\]

The second term is of order $O_{p}(n^{-1}\log n)$ (because $\hat{\theta}_{\mathrm{OLS}}-\theta=O_{p}(n^{-1/2})$
and Lemma \ref{lem: t1_uni} (i)). By similar arguments in p.22 of
BGH-supp and in the proof of Lemma \ref{lem: t1_uni} (i), the first
term is of order $O_{p}\left(n^{-1}(\log n)^{2}\right)$.

Then 
\begin{equation}
R_{n}=O_{p}\left(n^{-1}(\log n)^{2}\right).\label{pf:2-6}
\end{equation}
Thus, for some constants $C,K>0$ and a shrinking sequence $\epsilon_{n}$,
set inclusion relationships yield 
\begin{eqnarray*}
P(d_{2}^{2}(\hat{\tau},\tau_{0})\ge\epsilon_{n}^{2}) & = & P\left(d_{2}(\hat{\tau},\tau_{0})\ge\epsilon_{n},\text{ }\int g_{\hat{\tau}}(u,x)d(\mathbb{P}_{n}-P)(u,x)+R_{n}\ge d_{2}^{2}(\hat{\tau},\tau_{0})\right)\\
 & \le & P\left(\begin{array}{c}
d_{2}(\hat{\tau},\tau_{0})\ge\epsilon_{n},\text{ }|R_{n}|\le Cn^{-1}(\log n)^{2},\text{ }\left\Vert \hat{\tau}\right\Vert _{\infty}\le K\log n\\
\int g_{\hat{\tau}}(u,x)d(\mathbb{P}_{n}-P)(u,x)+R_{n}-d_{2}^{2}(\hat{\tau},\tau_{0})\ge0
\end{array}\right)\\
 &  & +P(|R_{n}|>Cn^{-1}(\log n)^{2})+P(\left\Vert \hat{\tau}\right\Vert _{\infty}>K\log n)\\
 & =: & P_{1}+P_{2}+P_{3},
\end{eqnarray*}
where the first equality follows from (\ref{pf:2-5}). For $P_{2}$
and $P_{3}$, (\ref{pf:2-6}) and Lemma \ref{lem: t1_uni} (i) imply
that we can choose $C$ and $K$ to make these terms arbitrarily small.
Thus, we focus on the first term $P_{1}$.

Now let 
\begin{eqnarray*}
\mathcal{T} & = & \{\tau:\tau\text{ is positive and monotone increasing on }\mathscr{X},\text{ }\left\Vert \tau\right\Vert _{\infty}\le K\log n\},\\
\mathcal{G} & = & \{g_{\tau}(u,x)=\{2u^{2}\tau(x)-\tau(x)^{2}\}-\{2u^{2}\tau_{0}(x)-\tau_{0}(x)^{2}\}:\text{ }\tau\in\mathcal{T}\},\\
\mathcal{G}_{v} & = & \{g\in\mathcal{G}:d_{2}(\tau,\tau_{0})\le v\}.
\end{eqnarray*}
Set inclusion relationships and Markov's inequality yield 
\begin{eqnarray*}
P_{1} & \le & P\left(\sup_{\tau\in\mathcal{T},d_{2}(\tau,\tau_{0})\ge\epsilon_{n}}\{\int g_{\tau}(u,x)d(\mathbb{P}_{n}-P)(u,x)-d_{2}^{2}(\tau,\tau_{0})\}\ge-Cn^{-1}(\log n)^{2}\right)\\
 & \le & \sum_{s=0}^{\infty}P\left(\sup_{\tau\in\mathcal{T},2^{s}\epsilon_{n}\le d_{2}(\tau,\tau_{0})\le2^{s+1}\epsilon_{n}}\sqrt{n}\{\int g_{\tau}(u,x)d(\mathbb{P}_{n}-P)(u,x)\}\ge\sqrt{n}\left(2^{2s}\epsilon_{n}^{2}-Cn^{-1}(\log n)^{2}\right)\right)\\
 & \le & \sum_{s=0}^{\infty}P\left(\left\Vert \mathbb{G}_{n}g\right\Vert _{\mathcal{G}_{2^{s+1}\epsilon_{n}}}\ge\sqrt{n}\left(2^{2s}\epsilon_{n}^{2}-Cn^{-1}(\log n)^{2}\right)\right)\\
 & \le & \sum_{s=0}^{\infty}E[\left\Vert \mathbb{G}_{n}g\right\Vert _{\mathcal{G}_{2^{s+1}\epsilon_{n}}}]/\{\sqrt{n}\left(2^{2s}\epsilon_{n}^{2}-Cn^{-1}(\log n)^{2}\right)\}.
\end{eqnarray*}
For a sufficiently large constant $\tilde{C}>0$, the sequence $\epsilon_{n}^{2}:=\tilde{C}(\log n)^{2}n^{-\frac{2}{3}}$
dominates $Cn^{-1}(\log n)^{2}$, so it holds $\sqrt{n}\left(2^{2s}\epsilon_{n}^{2}-Cn^{-1}(\log n)^{2}\right)=\sqrt{n}2^{2s}\epsilon_{n}^{2}(1+o(1))$.
Therefore, the standard result for the $L^{2}$-convergence of the
isotonic estimator under Assumption A2 (e.g., pp. 8-11 in BGH-supp)
implies that the last term can be made arbitrarily small by appropriately
selecting $\tilde{C}$. Thus, the proof is concluded.

\subsubsection{Proof of Lemma \ref{lem: t1_uni} (iii)}

We show $E[\left\Vert \mathbb{G}_{n}\right\Vert _{\mathcal{F}_{n}}]\le\frac{A\nu}{2}$
by using van der Vaart and Wellner (1996, Lemma 3.4.3). First we introduce
some notation for this part. Let $N_{[]}(\varepsilon,\mathcal{F},||\cdot||)$
be the $\varepsilon$-bracketing number of the function class $\mathcal{F}$
under the norm $||\cdot||$, $H_{B}(\varepsilon,\mathcal{F},||\cdot||)=\text{log}N_{[]}(\varepsilon,\mathcal{F},||\cdot||)$
be the entropy, $J_{n}(\delta,\mathcal{F},||\cdot||)=\int_{0}^{\delta}\sqrt{1+H_{B}(\varepsilon,\mathcal{F},||\cdot||)}d\varepsilon$,
and $\|f\|_{B,P}=(2E[e^{|f|}-|f|-1])^{1/2}$ be the Bernstein norm.

\textbf{Lemma 3.4.3 in van der Vaart and Wellner (1996): }Let $\mathcal{F}$
be a class of measurable functions such that $\|f\|_{B,P}^{2}\leq\delta$
for every $f$ in $\mathcal{F}$. Then 
\[
E[\left\Vert \mathbb{G}_{n}\right\Vert _{\mathcal{F}}]\lesssim J_{n}(\delta,\mathcal{F},||\cdot||_{B,P})\{1+J_{n}(\delta,\mathcal{F},||\cdot||_{B,P})/(\sqrt{n}\delta^{2})\}.
\]

To apply this lemma, we need to compute $H_{B}(\epsilon,\tilde{\mathcal{F}_{n}},\|\cdot\|_{B,P})$
and $\|\tilde{f}\|_{B,P}^{2}$, where $\tilde{\mathcal{F}_{n}}=\{\tilde{f}=D^{-1}f:f\in\mathcal{F}_{n}\}$,
the function class $\mathcal{F}_{n}$ is defined below in \eqref{pf:F_uni},
and the constant $D>0$ will be chosen later to guarantee that the
Bernstein norm of $\tilde{f}$ is finite. Moreover, let us define
the following function class:
\[
\mathcal{T}_{\mathcal{I},K_{1}}=\{\tau\mbox{ monotone non-decreasing on the interval }\mathcal{I}\mbox{ and }0<\tau<K_{1}\}.
\]
Assumption A2 implies that there exist positive constants, $\underline{C}$
and $\overline{C}$, such that $0<\underline{C}<\tau_{0}<\overline{C}<\infty$.
Also let 
\begin{equation}
\mathcal{F}_{n}=\left\{ f_{n}(w,u)=\mathbb{I}\{x>q_{n}\}\left(\frac{1}{\tau(x)}-\frac{1}{\tau_{0}(x)}\right)w_{h}u:\begin{array}{c}
\text{ }\tau\in\mathcal{T}_{\mathscr{X},K_{1}},\text{ }\left\Vert \tau-\tau_{0}\right\Vert _{2,P}^{2}\le v^{2},\\
\mathbb{I}\{x>q_{n}\}/\tau(x)\leq1/K_{0},h\in\{1:\mathrm{dim}(w)\}
\end{array}\text{ }\right\} ,\label{pf:F_uni}
\end{equation}
where $w_{h}$ is the $h$-th component of vector $w$. We set $2K_{0}=\underline{C}$,
$K_{1}=K_{2}\log n$, and $v=K_{3}(\log n)n^{-1/3}$ for some constants
$K_{2},K_{3}>0$.

Consider $\epsilon$-brackets $(\tau^{L},\tau^{U})$ under the $L_{2}(P)$-norm
for the functions in $\mathcal{T}_{\mathcal{I},K_{1}}$. According
to van der Vaart and Wellner (1996, Theorem 2.7.5), there exists some
constant $C^{\prime}>0$ such that 
\begin{equation}
H_{B}(\epsilon,\mathcal{T}_{\mathscr{X},K_{1}},\|\cdot\|_{2,P})\le\frac{C^{\prime}K_{1}}{\epsilon},\quad\text{for each }\epsilon\in(0,K_{1}).\label{pf:mono_bracket}
\end{equation}

Without loss of generality, we can choose those bracket functions
that satisfy $\mathbb{I}\{x>q_{n}\}/\tau^{L}(x)\leq1/K_{0}$.\footnote{By definition (\ref{pf:F_uni}), the $\tau(\cdot)$ associated to
$\mathcal{F}_{n}$ must satisfy $\mathbb{I}\{x>q_{n}\}/\tau(x)\leq1/K_{0}$.
Since $\mathcal{T}_{\mathscr{X},K_{1}}$ is a class of monotone increasing
function, any $\epsilon$-brackets of $\mathcal{T}_{\mathscr{X},K_{1}}$
can be modified to be a $\epsilon$-bracket of the ``$\mathcal{F}_{n}$-subset''
of $\mathcal{T}_{\mathscr{X},K_{1}}$, satisfying $\mathbb{I}\{x>q_{n}\}/\tau(x)\leq1/K_{0}$
by leveling-up certain part of lower bounds functions $\tau^{L}$,
without changing the bracket numbers, and the size of each modified
bracket can only be smaller.} Define 
\begin{eqnarray*}
f^{L}(w,u) & = & \begin{cases}
\mathbb{I}\{x>q_{n}\}\left(\frac{1}{\tau^{U}(x)}-\frac{1}{\tau_{0}(x)}\right)w_{h}u & \mbox{if }w_{h}u\ge0,\\
\mathbb{I}\{x>q_{n}\}\left(\frac{1}{\tau^{L}(x)}-\frac{1}{\tau_{0}(x)}\right)w_{h}u & \mbox{if }w_{h}u<0,
\end{cases}\\
f^{U}(w,u) & = & \begin{cases}
\mathbb{I}\{x>q_{n}\}\left(\frac{1}{\tau^{L}(x)}-\frac{1}{\tau_{0}(x)}\right)w_{h}u & \mbox{if }w_{h}u\ge0,\\
\mathbb{I}\{x>q_{n}\}\left(\frac{1}{\tau^{U}(x)}-\frac{1}{\tau_{0}(x)}\right)w_{h}u & \mbox{if }w_{h}u<0.
\end{cases}
\end{eqnarray*}
Note that $(f^{L},f^{U})$ is a bracket of $f\in\mathcal{F}_{n}$
for every $q_{n}\in[x_{L},x_{U}]$.

Now we compute the bracket size of $(\tilde{f}^{L},\tilde{f}^{U}):=(D^{-1}f^{L},D^{-1}f^{U})$
with respect to the Bernstein norm. Note that 
\begin{eqnarray*}
 &  & \|\tilde{f}^{U}-\tilde{f}^{L}\|_{B,P}^{2}=\|D^{-1}f^{U}-D^{-1}f^{L}\|_{B,P}^{2}\\
 & \leq & 2\sum_{k=2}^{\infty}\frac{1}{k!D^{k}}\int_{\mathcal{\mathscr{W}}\times\mathbb{R}}\left|\frac{\tau^{U}(x)-\tau^{L}(x)}{\tau^{L}(x)\tau^{U}(x)}w_{h}u\right|^{k}dP(w,u)\\
 & \le & 2\sum_{k=2}^{\infty}\frac{1}{k!D^{k}}\left\{ \frac{R^{k}k!M_{0}^{k-2}a_{0}(2K_{1})^{k-2}}{K_{0}^{2k}}\|\tau^{U}-\tau^{L}\|_{2,P}^{2}\right\} \le2a_{0}\left(\frac{R}{DK_{0}^{2}}\right)^{2}\sum_{k=0}^{\infty}\left(\frac{2RM_{0}K_{1}}{DK_{0}^{2}}\right)^{k}\epsilon^{2},
\end{eqnarray*}
where the first inequality follows from the definition of $\|\cdot\|_{B,P}^{2}$
and $\mathbb{I}\{x>q_{n}\}\le1$, the second inequality follows from
Assumption A2 (where we can choose $a_{0},M_{0}>1$) and $\frac{\mathbb{I}\{x>q_{n}\}}{\tau^{L}(x)}\leq\frac{1}{K_{0}}$.
Thus, by setting $D=4M_{0}RK_{1}/K_{0}^{2}$, we obtain$\|\tilde{f}^{U}-\tilde{f}^{L}\|_{B,P}^{2}\le\frac{a_{0}}{4M_{0}^{2}K_{1}^{2}}\epsilon^{2}$,
which implies 
\begin{align*}
\|\tilde{f}^{U}-\tilde{f}^{L}\|_{B,P} & \le\tilde{K}\epsilon,
\end{align*}
for $\tilde{K}=\frac{a_{0}^{1/2}}{2M_{0}K_{1}}$. Note that $(\tilde{f}^{L},\tilde{f}^{U})$
is: (a) a set of brackets in $\tilde{\mathcal{F}}_{n}$, (b) one-to-one
induced by ($\tau^{L},\tau^{U})$, an $\epsilon$-bracket in $\mathcal{T}_{\mathscr{X},K_{1}}$
with the entropy $H_{B}(\epsilon,\mathcal{T}_{\mathscr{X},K_{1}},\|\cdot\|_{2,P})$,
and (c) $\|\tilde{f}^{U}-\tilde{f}^{L}\|_{B,P}\leq\tilde{K}\epsilon$.
Based on these facts, (\ref{pf:mono_bracket}) yields 
\[
H_{B}(\tilde{K}\epsilon,\tilde{\mathcal{F}}_{n},\|\cdot\|_{B,P})\leq H_{B}(\epsilon,\mathcal{T}_{\mathscr{X},K_{1}},\|\cdot\|_{2,P})\leq\frac{C^{\prime}K_{1}}{\epsilon},
\]
which implies (by a change-of-variable argument) 
\begin{equation}
H_{B}(\epsilon,\tilde{\mathcal{F}}_{n},\|\cdot\|_{B,P})\le\frac{\tilde{K}C^{\prime}K_{1}}{\epsilon}=\frac{\tilde{B}}{\epsilon},\quad\text{for }\tilde{B}=\frac{C^{\prime}a_{0}^{1/2}}{2M_{0}}.\label{pf:bracket_n_uni}
\end{equation}

We now characterize the Bernstein norm of $\tilde{f}$, 
\begin{eqnarray*}
\|\tilde{f}\|_{B,P}^{2} & \leq & 2\sum_{k=2}^{\infty}\frac{1}{k!D^{k}}\int_{\mathcal{\mathscr{W}}\times\mathbb{R}}\left|\frac{\tau(x)-\tau_{0}(x)}{\tau(x)\tau_{0}(x)}w_{h}u\right|^{k}dP(w,u)\\
 & \le & 2\sum_{k=2}^{\infty}\frac{1}{k!D^{k}}\left\{ \frac{R^{k}k!M_{0}^{k-2}a_{0}(2K_{1})^{k-2}}{K_{0}^{2k}}\|\tau-\tau_{0}\|_{2,P}^{2}\right\} \\
 & \le & 2a_{0}\left(\frac{R}{DK_{0}^{2}}\right)^{2}\sum_{k=0}^{\infty}\left(\frac{2RM_{0}K_{1}}{DK_{0}^{2}}\right)^{k}v^{2}\le\frac{a_{0}}{4M_{0}^{2}}\frac{1}{K_{1}^{2}}v^{2},
\end{eqnarray*}
where the second inequality follows from $\frac{\mathbb{I}\{x>q_{n}\}}{\tau(x)}\leq\frac{1}{K_{0}}$,
and the third inequality follows from \eqref{pf:F_uni} and some rearrangements.
Then, we have 
\begin{equation}
\|\tilde{f}\|_{B,P}\le\frac{Bv}{K_{1}},\quad\text{for }B=\frac{a_{0}^{1/2}}{2M_{0}}.\label{pf:berstein_norm_uni}
\end{equation}

Combining (\ref{pf:bracket_n_uni}) and (\ref{pf:berstein_norm_uni}),
van der Vaart and Wellner (1996, Lemma 3.4.3) implies 
\begin{align*}
E[\left\Vert \mathbb{G}_{n}\right\Vert _{\tilde{\mathcal{F}}_{n}}] & \lesssim J_{n}(BK_{1}^{-1}v)\left(1+\frac{J_{n}(BK_{1}^{-1}v)}{\sqrt{n}B^{2}v^{2}/K_{1}^{2}}\right),
\end{align*}
where $J_{n}(\cdot)$ is the abbreviation of $J_{n}(\cdot,\tilde{\mathcal{F}}_{n},\|\cdot\|_{B,P})$.
By the arguments used in the proof of Proposition 7.9 of BDJ, it holds
\[
J_{n}(BK_{1}^{-1}v)\le BK_{1}^{-1}v+2\tilde{B}^{1/2}B^{1/2}K_{1}^{-1/2}v^{1/2}\lesssim B_{1}K_{1}^{-1/2}v^{1/2},
\]
for some $B_{1}>0$ and sufficiently small $v$. This implies 
\[
E[\left\Vert \mathbb{G}_{n}\right\Vert _{\tilde{\mathcal{F}_{n}}}]\lesssim B_{1}K_{1}^{-1/2}v^{1/2}\left(1+K_{1}^{2}\frac{B_{1}K_{1}^{-1/2}v^{1/2}}{\sqrt{n}B^{2}v^{2}}\right)\lesssim B_{1}K_{1}^{-1/2}v^{1/2}\left(1+\frac{B_{2}K_{1}^{3/2}}{\sqrt{n}v^{3/2}}\right),
\]
for some $B_{2}>0$. By the definition of the class $\tilde{\mathcal{F}}_{n}=\{\tilde{f}=D^{-1}f:f\in\mathcal{F}_{n}\},$
it follows that 
\[
E[\left\Vert \mathbb{G}_{n}\right\Vert _{\mathcal{F}_{n}}]=D\cdot E[\left\Vert \mathbb{G}_{n}\right\Vert _{\tilde{\mathcal{F}}_{n}}]\lesssim DB_{1}K_{1}^{-1/2}v^{1/2}\left(1+\frac{B_{2}K_{1}^{3/2}}{\sqrt{n}v^{3/2}}\right)\lesssim B_{3}K_{0}^{-2}K_{1}^{1/2}v^{1/2}\left(1+\frac{B_{2}K_{1}^{3/2}}{\sqrt{n}v^{3/2}}\right),
\]
for some $B_{3}>0$. The conclusion follows by observing that with
$v=K_{3}(\log n)n^{-1/3}$, $K_{1}=K_{2}\log n$, and all sufficiently
large $n$, we have
\[
E[\left\Vert \mathbb{G}_{n}\right\Vert _{\mathcal{F}_{n}}]\lesssim C_{3}(\log n)n^{-1/6}(1+C_{4})\lesssim\frac{A\nu}{2},
\]
where $C_{3}=B_{3}K_{0}^{-2}K_{2}^{1/2}K_{3}^{1/2}$ and $C_{4}=B_{2}(K_{2}/K_{3})^{3/2}$. 

\section{Proof of lemma and theorem in Section \ref{sec:case2}\label{app:multi-proofs}}

\textbf{Notation.} To avoid heavy notations, some of them are used
in Appendix \ref{app:uni-proofs} but redefined here. Define $\tau_{\eta}(a)=E[\sigma^{2}(X^{\prime}\eta_{0})|X^{\prime}\eta=a]$
and $\tau_{\eta_{0}}(a)=\tau_{0}(a)$ (note that $\tau_{0}(x^{\prime}\eta_{0})=\sigma^{2}(x^{\prime}\eta_{0})$).
Let $\hat{\tau}_{\eta}=\hat{\tau}_{\eta}(x^{\prime}\eta)$ be the
isotonic estimator obtained by (\ref{eq:p_gamma}) for a given $\eta$,
$\mathscr{W}$ be the support of $W:=(1,X^{\prime},Z^{\prime})^{\prime}$,
$F_{n}(t)=\frac{1}{n}\sum_{i=1}^{n}\mathbb{I}\{X_{i}^{\prime}\hat{\eta}\le t\}$,
and $M_{n}(t)=\frac{1}{n}\sum_{i=1}^{n}\hat{U}_{i}^{2}\mathbb{I}\{X_{i}^{\prime}\hat{\eta}\le t\}$.

\subsection{Proof of Lemma \ref{lem:trun_multi}\label{app-sub:Proof lemma 1}}

The main part of the proof is similar to that of Lemma \ref{lem:trun_uni}.
Recall that $q_{n}^{*}$ is the $(n^{-1/3})$-th population quantile
of $(X^{\prime}\eta_{0})$ and $q_{n}$ is the $(n^{-1/3})$-th sample
quantile of $\{X_{i}^{\prime}\hat{\eta}\}_{i=1}^{n}$ with $\hat{\eta}$
estimated by (\ref{eq:simplescore}). To proceed, we use the following
lemma:

\begin{lem} \label{lem:BGH_e_hat} Under Assumptions M1-M6, it holds 
\begin{description}
\item [{(i)}] $\hat{\eta}-\eta_{0}=O_{p}(n^{-1/2})$, 
\item [{(ii)}] $\tau_{\hat{\eta}}(a)-\tau_{0}(a)=O_{p}(n^{-1/2})$ for
each $a$, and $\left\Vert \tau_{\hat{\eta}}-\tau_{0}\right\Vert _{2,P}=O_{p}(n^{-1/2})$. 
\end{description}
\end{lem}

The proof of this lemma is in Appendix \ref{app-sub:e_BGH_proof}.
Based on Lemma \ref{lem:BGH_e_hat} (i), Assumptions M2-M3, and properties
of the sample quantile, we obtain $q_{n}-q_{n}^{*}=O_{p}(n^{-1/2})=o_{p}(n^{-1/3})$,
which implies $c^{*}=\lim_{n\to\infty}n^{1/3}(q_{n}^{*}-x_{L})=\text{plim}_{n\to\infty}n^{1/3}(q_{n}-x_{L})<\infty$.
By Assumption M2, Lemma \ref{lem:BGH_e_hat} (ii), and similar arguments
in Appendix \ref{app-sub:lem1}, we have 
\begin{eqnarray*}
 &  & n^{1/3}\{\hat{\tau}_{\hat{\eta}}(q_{n})-\tau_{0}(q_{n})\}=n^{1/3}\{\hat{\tau}_{\hat{\eta}}(q_{n})-\tau_{\hat{\eta}}(q_{n})\}+o_{p}(1)\\
 & = & n^{1/3}[\{\hat{\tau}_{\hat{\eta}}(q_{n})-\tau_{\hat{\eta}}(x_{L})\}-\{\tau_{\hat{\eta}}(q_{n})-\tau_{0}(x_{L})\}]+o_{p}(1)\\
 & \overset{d}{\to} & D_{[0,\infty)}^{L}\left(\sqrt{\frac{\sigma_{\varepsilon}^{2}(x_{L})}{c^{*}f_{X}(x_{L})}}\mathcal{W}_{t}+\tau_{0}^{\prime}(x_{L})\frac{t^{2}c^{*}}{2}\right)(1)-\underset{n\to\infty}{\text{plim}}n^{1/3}\{\tau_{0}(q_{n})-\tau_{0}(x_{L})\}\\
 & \overset{d}{\sim} & D_{[0,\infty)}^{L}\left(\sqrt{\frac{\sigma_{\varepsilon}^{2}(x_{L})}{c^{*}f_{X}(x_{L})}}\mathcal{W}_{t}+\tau_{0}^{\prime}(x_{L})\frac{t^{2}c^{*}}{2}\right)(1)-\underset{n\to\infty}{\text{lim}}n^{1/3}\{\tau_{0}(q_{n}^{*})-\tau_{0}(x_{L})\}\\
 & \overset{d}{\sim} & D_{[0,\infty)}^{L}\left(\sqrt{\frac{\sigma_{\varepsilon}^{2}(x_{L})}{c^{*}f_{X}(x_{L})}}\mathcal{W}_{t}+\tau_{0}^{\prime}(x_{L})\frac{t^{2}c^{*}}{2}-\tau_{0}^{\prime}(x_{L})c^{*}t\right)(1),
\end{eqnarray*}
where the first and second equalities follow from Lemma \ref{lem:BGH_e_hat}
(ii), the convergence follows from a similar argument to (\ref{pf:concl}),
the first distribution relation follows from Lemma \ref{lem:BGH_e_hat}
(ii), Assumption M2(iv), and $q_{n}^{*}-q_{n}=o_{p}(n^{-1/3})$, and
the second distribution relation follows from the fact that the $D_{[0,\infty)}^{L}$
is a linear operator for a linear function of $t$.

\subsection{Proof of Theorem \ref{thm:multi}}

Similar to Theorem \ref{thm:uni}, it is sufficient for the conclusion
to prove the following lemma.

\begin{lem} \label{lem: t1_multi} Under Assumptions M1-M6, it holds 
\begin{description}
\item [{(i)}] $\left\Vert \hat{\tau}_{\eta}\right\Vert _{\infty}=O_{p}(\log n)$
uniformly over $\eta\in\mathcal{B}(\eta_{0},\delta_{0})$, 
\item [{(ii)}] $\left\Vert \hat{\tau}_{\hat{\eta}}-\tau_{0}\right\Vert _{2,P}^{2}=O_{p}((\log n)^{2}n^{-2/3})$, 
\item [{(iii)}] $E\left[\left\Vert \mathbb{G}_{n}\right\Vert _{\mathcal{F}_{n}}\right]\le\frac{A\nu}{2}$
holds for any constants $A>0$ and $\nu>0$, and all sufficiently
large $n$, where $\mathcal{F}_{n}$ is the function class defined
as
\[
\mathcal{F}_{n}=\left\{ f_{n}(w,u)=\mathbb{I}\{x^{\prime}\eta>q_{n}\}\left(\frac{1}{\tau(x^{\prime}\eta)}-\frac{1}{\tau_{\eta}(x^{\prime}\eta)}\right)w_{h}u:\begin{array}{c}
\tau\ge0\text{ is monotone increasing on }I_{\eta},\\
\left\Vert \tau\right\Vert _{\infty}\le C\log n,\text{ }\left\Vert \tau-\tau_{\eta}\right\Vert _{2,P}^{2}\le Cr_{n},\\
\mathbb{I}(x^{\prime}\eta>q_{n})/\tau(x^{\prime}\eta)\leq1/K_{0},\\
h\in\{1:\mathrm{dim}(w)\}
\end{array}\text{ }\right\} ,
\]
with $C$ and $K_{0}$ being some positive constants, and $r_{n}=(\log n)^{2}n^{-2/3}$.
\end{description}
\end{lem}

\subsubsection{Proof of Lemma \ref{lem: t1_multi} (i)}

The proof is adapted from BDJ (2019, eq. (7.11) on p.3297). For fixed
$\eta$, let $\{\hat{U}_{\eta,i}^{2}\}_{i=1}^{n}$ be a permutation
of $\{\hat{U}_{j}^{2}\}_{j=1}^{n}$, which is arranged according to
the monotonically ordered series $\{X_{i}^{\prime}\eta\}_{i=1}^{n}$.
The min-max formula of the isotonic regression says 
\[
\min_{1\le k\le n}\frac{\sum_{i=1}^{k}\hat{U}_{\eta,i}^{2}}{k}\le\hat{\tau}_{\eta}(x^{\prime}\eta)\le\max_{1\le k\le n}\frac{\sum_{i=k}^{n}\hat{U}_{\eta,i}^{2}}{n-k+1},
\]
for each $x\in\mathscr{X}$ and $\eta\in\mathcal{B}(\eta_{0},\delta_{0})$,
which implies $\min_{1\le j\le n}\hat{U}_{j}^{2}\le\hat{\tau}_{\eta}(x^{\prime}\eta)\le\max_{1\le j\le n}\hat{U}_{j}^{2}$
for each $x\in\mathscr{X}$. Thus, it is sufficient for the conclusion
to show that 
\begin{equation}
\max_{1\le j\le n}\hat{U}_{j}^{2}=O_{p}(\log n).\label{pf:1-1-1}
\end{equation}
Observe that 
\[
\max_{1\le j\le n}\hat{U}_{j}^{2}\le\max_{1\le j\le n}U_{j}^{2}+2Rk||\hat{\theta}_{\mathrm{OLS}}-\theta||_{\infty}\max_{1\le j\le n}|U_{j}|+R^{2}k^{2}||\hat{\theta}_{\mathrm{OLS}}-\theta||_{\infty}^{2},
\]
where $k$ is the dimension of $\theta$. From Lemma 7.1 of BDJ, Assumption
M2 guarantees $\max_{1\le j\le n}U_{j}^{2}=O_{p}(\log n)$. By the
same reasoning for the proof of Lemma \ref{lem: t1_uni}, we have
$\max_{1\le j\le n}|U_{j}|=O_{p}(\log n)$ and $||\hat{\theta}_{\mathrm{OLS}}-\theta||_{\infty}=O_{p}(n^{-1/2})$.
Thus, we have $\left\Vert \hat{\tau}_{\eta}\right\Vert _{\infty}=O_{p}(\log n)$.
Since different $\eta$ only changes the permutation $\{\hat{U}_{\eta,i}^{2}\}_{i=1}^{n}$
but not $\max_{1\le j\le n}\hat{U}_{j}^{2}$, we have $\left\Vert \hat{\tau}_{\eta}\right\Vert _{\infty}=O_{p}(\log n)$
uniformly over $\eta\in\mathcal{B}(\eta_{0},\delta_{0}).$

\subsubsection{Proof of Lemma \ref{lem: t1_multi} (ii)}

The main part of the proof is similar to those of Lemma \ref{lem: t1_uni}
(ii) and Proposition 4 of BGH-supp. Define 
\begin{eqnarray*}
g_{\eta,\tau}(u,x) & = & \{2u^{2}\tau(x^{\prime}\eta)-\tau(x^{\prime}\eta)^{2}\}-\{2u^{2}\tau_{\eta}(x^{\prime}\eta)-\tau_{\eta}(x^{\prime}\eta)^{2}\},\\
R_{n,\eta} & = & \frac{2}{n}\sum_{j=1}^{n}\left(-2U_{j}W_{j}(\hat{\theta}_{\mathrm{OLS}}-\theta)+\{W_{j}(\hat{\theta}_{\mathrm{OLS}}-\theta)\}^{2}\right)\{\hat{\tau}_{\eta}(X_{j}^{\prime}\eta)-\tau_{\eta}(X_{j}^{\prime}\eta)\},\\
d_{2}^{2}(\tau_{1},\tau_{2}) & = & -E[2\tau_{1}\tau_{2}-\tau_{1}^{2}-\tau_{2}^{2}],
\end{eqnarray*}
Following reasoning similar to that presented for (\ref{pf:2-1})-(\ref{pf:2-6}),
we have for some $C$ and $K$, 
\begin{eqnarray*}
 &  & P(\sup_{\eta\in\mathcal{B}(\eta_{0},\delta_{0})}d_{2}^{2}(\hat{\tau}_{\eta},\tau_{\eta})\ge\epsilon_{n}^{2})\\
 & \le & P\left(\begin{array}{c}
\sup_{\eta\in\mathcal{B}(\eta_{0},\delta_{0})}d_{2}(\hat{\tau}_{\eta},\tau_{\eta})\ge\epsilon_{n},\text{ }\sup_{\eta\in\mathcal{B}(\eta_{0},\delta_{0})}\left\Vert \hat{\tau}_{\eta}\right\Vert _{\infty}\le K\log n,\\
\sup_{\eta\in\mathcal{B}(\eta_{0},\delta_{0})}\int g_{\eta,\hat{\tau}}(u,x)d(\mathbb{P}_{n}-P)(u,x)+R_{n,\eta}-d_{2}^{2}(\hat{\tau}_{\eta},\tau_{\eta})\ge0,\\
\sup_{\eta\in\mathcal{B}(\eta_{0},\delta_{0})}|R_{n,\eta}|\le Cn^{-1}(\log n)^{2}
\end{array}\right)\\
 &  & +P(|R_{n,\eta}|>Cn^{-1}(\log n)^{2})+P\left(\sup_{\eta\in\mathcal{B}(\eta_{0},\delta_{0})}\left\Vert \hat{\tau}_{\eta}\right\Vert _{\infty}>K\log n\right)\\
 & =: & P_{1}+P_{2}+P_{3}.
\end{eqnarray*}

Lemma \ref{lem: t1_multi} (i) implies $P_{3}\to0$, and $P_{2}\to0$
follows from similar arguments for (\ref{pf:2-6}). For $P_{1}$,
we define 
\begin{eqnarray*}
\mathcal{T} & = & \{\tau:\tau\text{ is positive and monotone increasing function on }I_{\eta},\text{ }\left\Vert \tau\right\Vert _{\infty}\le K\log n\},\\
\mathcal{G} & = & \{g(x,u)=\{2u^{2}\tau(x^{\prime}\eta)-\tau(x^{\prime}\eta)^{2}\}-\{2u^{2}\tau_{\eta}(x^{\prime}\eta)-\tau_{\eta}(x^{\prime}\eta)^{2}\}:\text{ }\tau\in\mathcal{T}\},\\
\mathcal{G}_{v} & = & \{g\in\mathcal{G}:d_{2}(\tau,\tau_{\eta})\le v\},
\end{eqnarray*}
for each $\eta\in\mathcal{B}(\eta_{0},\delta_{0})$. By similar arguments
for Lemma \ref{lem: t1_uni} (ii) and Proposition 4 of BGH-supp, we
can obtain
\begin{eqnarray*}
P_{1} & \le & \sum_{s=0}^{\infty}E\left[\left\Vert \mathbb{G}_{n}g\right\Vert _{\mathcal{G}_{2^{s+1}\epsilon_{n}}}\right]/\{\sqrt{n}2^{2s}\epsilon_{n}^{2}-Cn^{-1/2}(\log n)^{2}\},
\end{eqnarray*}
and
\begin{equation}
\sup_{\eta\in\mathcal{B}(\eta_{0},\delta_{0})}\int\{\hat{\tau}_{\eta}(x^{\prime}\eta)-\tau_{\eta}(x^{\prime}\eta)\}^{2}dF(x)=O_{p}((\log n)^{2}n^{-2/3}).\label{pf:BGH6.2}
\end{equation}
By combining \eqref{pf:BGH6.2}, Lemma \ref{lem:BGH_e_hat}, and the
triangle inequality, we obtain $\left\Vert \hat{\tau}_{\hat{\eta}}-\tau_{0}\right\Vert _{2,P}^{2}=O_{p}((\log n)^{2}n^{-2/3})$.

\subsubsection*{Proof of Lemma \ref{lem: t1_multi} (iii)}

To avoid heavy notation, we use the same notation as in the proof
of Lemma \ref{lem: t1_uni} (iii), but some notation is redefined
here. Let 
\[
\mathcal{T}_{\mathcal{I},K_{1}}=\{\tau\mbox{ monotone non-decreasing on some interval }\mathcal{I}\mbox{ and }0<\tau<K_{1}\}.
\]
Assumption M2 guarantees $0<\underline{C}<\tau_{0}<\overline{C}<\infty$.
Similar to the proof of Lemma \ref{lem: t1_uni} (iii), we calculate
$H_{B}(\epsilon,\tilde{\mathcal{F}},\|\cdot\|_{B,P})$ and $\|\tilde{f}\|_{B,P}^{2}$,
with $\tilde{\mathcal{F}}=\{\tilde{f}=D^{-1}f:f\in\mathcal{F}\}$,
where the constant $D>0$ is determined later. Define $I_{\eta}^{*}=(a^{L},a^{U})$
with $a^{L}=\inf_{x\in\mathscr{X},\eta\in\mathcal{B}(\eta_{0},\delta_{0})}x^{\prime}\eta$
and $a^{U}=\sup_{x\in\mathscr{X},\eta\in\mathcal{B}(\eta_{0},\delta_{0})}x^{\prime}\eta$.
Define 
\[
\mathcal{F}_{n}=\left\{ f_{n}(w,u)=\mathbb{I}\{x^{\prime}\eta>q_{n}\}\left(\frac{1}{\tau(x^{\prime}\eta)}-\frac{1}{\tau_{\eta}(x^{\prime}\eta)}\right)w_{h}u:\begin{array}{c}
\tau\in\mathcal{T}_{I_{\eta}^{*},K_{1}},\eta\in\mathcal{B}(\eta_{0},\delta_{0}),\\
\text{ }\left\Vert \tau-\tau_{0}\right\Vert _{2,P}^{2}\le v^{2},h\in\{1:\mathrm{dim}(w)\},\\
\mathbb{I}(x^{\prime}\eta>q_{n})/\tau(x)\leq1/K_{0}
\end{array}\right\} ,
\]
where $w_{h}$ is the $h$-th component of $w$. We set $2K_{0}=\underline{C}$,
$K_{1}=K_{2}\log n$, and $v=K_{3}(\log n)n^{-1/3}$ for some positive
constants $K_{2}$ and $K_{3}$.

By van der Vaart and Wellner (1996, Theorem 2.7.5), it holds for each
$\epsilon\in(0,K_{1})$, 
\[
H_{B}(\epsilon,\mathcal{T}_{I_{\eta}^{*},K_{1}},\|\cdot\|_{P})\le\frac{C^{\prime}K_{1}}{\epsilon}.
\]
Similarly to the univariate case, we can choose those bracket functions
$(\tau^{L},\tau^{U})$, which satisfy $\mathbb{I}\{x^{\prime}\eta>q_{n}\}/\tau^{L}(x^{\prime}\eta)\leq1/K_{0}$.
Then, we define 
\begin{eqnarray*}
f^{L}(w,u) & = & \begin{cases}
\mathbb{I}\{x^{\prime}\eta>q_{n}\}\left(\frac{1}{\tau^{U}(x^{\prime}\eta)}-\frac{1}{\tau_{\eta}(x^{\prime}\eta)}\right)w_{h}u & \mbox{if }w_{h}u\ge0,\\
\mathbb{I}\{x^{\prime}\eta>q_{n}\}\left(\frac{1}{\tau^{L}(x^{\prime}\eta)}-\frac{1}{\tau_{\eta}(x^{\prime}\eta)}\right)w_{h}u & \mbox{if }w_{h}u<0,
\end{cases}\\
f^{U}(w,u) & = & \begin{cases}
\mathbb{I}\{x^{\prime}\eta>q_{n}\}\left(\frac{1}{\tau^{L}(x^{\prime}\eta)}-\frac{1}{\tau_{\eta}(x^{\prime}\eta)}\right)w_{h}u & \mbox{if }w_{h}u\ge0,\\
\mathbb{I}\{x^{\prime}\eta>q_{n}\}\left(\frac{1}{\tau^{U}(x^{\prime}\eta)}-\frac{1}{\tau_{\eta}(x^{\prime}\eta)}\right)w_{h}u & \mbox{if }w_{h}u<0.
\end{cases}
\end{eqnarray*}
Note that $(f^{L},f^{U})$ is a bracket for $f\in\mathcal{F}_{n}$.
The bracket size is 
\begin{eqnarray*}
 &  & \|\tilde{f}^{U}-\tilde{f}^{L}\|_{B,P}^{2}=\|D^{-1}f^{U}-D^{-1}f^{L}\|_{B,P}^{2}\\
 & = & 2\sum_{k=2}^{\infty}\frac{1}{k!D^{k}}\int_{\mathcal{\mathscr{W}}\times\mathbb{R}}\mathbb{I}\{x^{\prime}\eta>q_{n}\}\left|\left(\frac{1}{\tau^{L}(x^{\prime}\eta)}-\frac{1}{\tau^{U}(x^{\prime}\eta)}\right)w_{h}u\right|^{k}dP(w,u)\\
 & \le & 2\sum_{k=2}^{\infty}\frac{1}{k!D^{k}}\left\{ \frac{R^{k}k!M_{0}^{k-2}a_{0}(2K_{1})^{k-2}}{K_{0}^{2k}}\|\tau^{U}-\tau^{L}\|_{P}^{2}\right\} \\
 & \le & 2a_{0}\left(\frac{R}{DK_{0}^{2}}\right)^{2}\sum_{k=0}^{\infty}\left(\frac{2RM_{0}K_{1}}{DK_{0}^{2}}\right)^{k}\epsilon^{2},
\end{eqnarray*}
where the first inequality follows from Assumption M2 (where we can
choose $a_{0},M_{0}>1$) and $\frac{\mathbb{I}\{x^{\prime}\eta>q_{n}\}}{\tau^{L}(x^{\prime}\eta)}\leq\frac{1}{K_{0}}$.
Setting $D=4M_{0}RK_{1}/K_{0}^{2}$ yields $\|\tilde{f}^{U}-\tilde{f}^{L}\|_{B,P}\le\tilde{K}\epsilon$
for $\tilde{K}=\frac{a_{0}^{1/2}}{2M_{0}K_{1}}$, and thus 
\begin{align}
H_{B}(\epsilon,\tilde{\mathcal{F}},\|\cdot\|_{B,P}) & \leq\frac{\tilde{B}}{\epsilon},\quad\text{for }\tilde{B}=\frac{C_{2}a_{0}^{1/2}}{2M_{0}}.\label{pf:Bracket_n_multi}
\end{align}

Now we compute the Bernstein norm of $\tilde{f}$: 
\begin{eqnarray*}
\|\tilde{f}\|_{B,P}^{2} & = & 2\sum_{k=2}^{\infty}\frac{1}{k!D^{k}}\int_{\mathcal{\mathscr{W}}\times\mathbb{R}}\mathbb{I}\{x^{\prime}\eta>q_{n}\}\left|\left(\frac{1}{\tau(x^{\prime}\eta)}-\frac{1}{\tau_{\eta}(x^{\prime}\eta)}\right)w_{h}u\right|^{k}dP(w,u)\\
 & \le & 2\sum_{k=2}^{\infty}\frac{1}{k!D^{k}}\left\{ \frac{R^{k}k!M_{0}^{k-2}a_{0}(2K_{1})^{k-2}}{K_{0}^{2k}}\|\tau-\tau_{0}\|_{P}^{2}\right\} \\
 & \le & 2a_{0}\left(\frac{R}{DK_{0}^{2}}\right)^{2}\sum_{k=0}^{\infty}\left(\frac{2RM_{0}K_{1}}{DK_{0}^{2}}\right)^{k}v^{2}\le\frac{a_{0}}{4M_{0}^{2}}\frac{1}{K_{1}^{2}}v^{2},
\end{eqnarray*}
where the first inequality follows from $\frac{\mathbb{I}\{x^{\prime}\eta>q_{n}\}}{\tau(x^{\prime}\eta)}\leq\frac{1}{K_{0}}$.
This implies 
\begin{equation}
\|\tilde{f}\|_{B,P}\le B\frac{v}{K_{1}},\quad\text{for }B=\frac{a_{0}^{1/2}}{2M_{0}}.\label{pf:Bernstein_norm_multi}
\end{equation}

Combining (\ref{pf:Bracket_n_multi}) and (\ref{pf:Bernstein_norm_multi}),
the remaining steps are the same as those in the proof of Lemma \ref{lem: t1_uni}
(iii).

\subsection{Proof of Lemma \ref{lem:BGH_e_hat}\label{app-sub:e_BGH_proof}}

Recall for fixed $\eta$, we first obtain $\hat{\tau}_{\eta}=\arg\min_{\tau\in\mathcal{M}}\frac{1}{n}\sum_{i=1}^{n}\{\hat{U}_{i}^{2}-\tau(X_{i}^{\prime}\eta)\}^{2}$
and then obtain $\hat{\eta}$ by $\hat{\eta}=\text{arg}\underset{\eta}{\text{min}}||\frac{1}{n}\sum_{i=1}^{n}X_{i}^{\prime}\{\hat{U}_{i}^{2}-\hat{\tau}_{\eta}(X_{i}^{\prime}\eta)\}||^{2}$.
We denote $E[X|X^{\prime}\eta=x^{\prime}\eta]$ by $E[X|x^{\prime}\eta]$.
The proof is similar to the ones in BGH and Balabdaoui and Groeneboom
(2021) except that we need to handle the influence of the estimated
dependent variables $\hat{U}_{i}^{2}$.

The proof of consistency of $\hat{\eta}$ is similar to pp.16-17 of
BGH-supp. By a similar argument in Balabdaoui and Groeneboom (2021,
Lemma 3.2), under Assumptions M1-M3, we have 
\[
\frac{1}{n}\sum_{i=1}^{n}X_{i}^{\prime}\{\hat{U}_{i}^{2}-\hat{\tau}_{\eta}(X_{i}^{\prime}\eta)\}=\frac{1}{n}\sum_{i=1}^{n}(X_{i}-E[X|X_{i}^{\prime}\eta])\{\hat{U}_{i}^{2}-\tau_{\eta}(X_{i}^{\prime}\eta)\}+o_{p}(n^{-1/2}),
\]
for each $\eta$, where we also use (\ref{pf:BGH6.2}). Thus, it holds
\begin{eqnarray*}
\left\Vert \frac{1}{n}\sum_{i=1}^{n}X_{i}\{\hat{U}_{i}^{2}-\hat{\tau}_{\hat{\eta}}(X_{i}^{\prime}\hat{\eta})\}\right\Vert  & = & \underset{\eta}{\text{min}}\left\Vert \frac{1}{n}\sum_{i=1}^{n}X_{i}\{\hat{U}_{i}^{2}-\hat{\tau}_{\eta}(X_{i}^{\prime}\eta)\}\right\Vert \\
 & \leq & \underset{\eta}{\text{min}}\left\Vert \frac{1}{n}\sum_{i=1}^{n}(X_{i}-E[X|X_{i}^{\prime}\eta])\{\hat{U}_{i}^{2}-\tau_{\eta}(X_{i}^{\prime}\eta)\}+o_{p}(n^{-1/2})\right\Vert .
\end{eqnarray*}
The leading term inside the norm $\left\Vert \cdot\right\Vert $ of
the last expression does not depend on the potentially non-smooth
$\hat{\tau}_{\eta}$; it is a smooth function of $\eta$. Thus, under
standard conditions for the method of moments, we have $\min_{\eta}\left\Vert \frac{1}{n}\sum_{i=1}^{n}(X_{i}-E[X_{i}|X_{i}^{\prime}\eta])\{\hat{U}_{i}^{2}-\tau_{\eta}(X_{i}^{\prime}\eta)\}\right\Vert =0$,
and
\begin{eqnarray}
o_{p}(n^{-1/2}) & = & \frac{1}{n}\sum_{i=1}^{n}X_{i}\{\hat{U}_{i}^{2}-\hat{\tau}_{\hat{\eta}}(X_{i}^{\prime}\hat{\eta})\}\nonumber \\
 & = & \frac{1}{n}\sum_{i=1}^{n}(X_{i}-E[X|X_{i}^{\prime}\hat{\eta}])\{\hat{U}_{i}^{2}-\hat{\tau}_{\hat{\eta}}(X_{i}^{\prime}\hat{\eta})\}+o_{p}(n^{-1/2}+(\hat{\eta}-\eta))\nonumber \\
 & = & \int(x-E[X|x^{\prime}\hat{\eta}])\{\hat{u}^{2}-\tau_{0}(x^{\prime}\eta_{0})\}d(\mathbb{P}_{n}-P)(x,\hat{u})\nonumber \\
 &  & +\int(x-E[X|x^{\prime}\hat{\eta}])\{\hat{u}^{2}-\tau_{\hat{\eta}}(x^{\prime}\hat{\eta})\}dP(x,\hat{u})+o_{p}(n^{-1/2}+(\hat{\eta}-\eta))\nonumber \\
 & =: & I+II+o_{p}(n^{-1/2}+(\hat{\eta}-\eta)),\label{pf:score_decompo}
\end{eqnarray}
where the second equality follows from similar arguments to pp.18-20
of BGH-supp and (\ref{pf:BGH6.2}), and the third equality follows
from a similar argument in pp.21-23 of BGH-supp.

Let $\hat{U}(w,u)=u-w^{\prime}(\hat{\theta}_{\mathrm{OLS}}-\theta)$
and 
\begin{equation}
\hat{e}(w,u):=\hat{U}(w,u)^{2}-u^{2}=-2w^{\prime}(\hat{\theta}_{\mathrm{OLS}}-\theta)u+\{w^{\prime}(\hat{\theta}_{\mathrm{OLS}}-\theta)\}^{2}.\label{pf:en}
\end{equation}
For $I$, we have
\begin{eqnarray}
I & = & \int(x-E[X|x^{\prime}\hat{\eta}])\{u^{2}+\hat{e}(w,u)-\tau_{0}(x^{\prime}\eta_{0})\}d(\mathbb{P}_{n}-P)(w,u)\nonumber \\
 & = & \int(x-E[X|x^{\prime}\eta_{0}])\{u^{2}-\tau_{0}(x^{\prime}\eta_{0})\}d(\mathbb{P}_{n}-P)(x,u)\nonumber \\
 &  & +\int(x-E[X|x^{\prime}\hat{\eta}])\hat{e}(w,u)d(\mathbb{P}_{n}-P)(w,u)+o_{p}(n^{-1/2})\nonumber \\
 & = & \int(x-E[X|x^{\prime}\eta_{0}])\{u^{2}-\tau_{0}(x^{\prime}\eta_{0})\}d(\mathbb{P}_{n}-P)(x,u)+o_{p}(n^{-1/2}),\label{pf:score_1}
\end{eqnarray}
where the second equality follows from p.21 of BGH-supp, and the third
equality follows from the facts that (a) $\hat{\theta}_{\mathrm{OLS}}-\theta=O_{p}(n^{-1/2})$,
(b) $\hat{e}(w,u)$ is a parametric function of $w$ and $u$ in a
changing class indexed by $\hat{\theta}_{\mathrm{OLS}}$ (see (\ref{pf:en})),
so its $\epsilon$-entropy is of order $\log(1/\epsilon)\le1/\epsilon$
(see, e.g., Example 19.7 of van der Vaart and Wellner, 2000), and
(c) similar arguments in pp.22-23 of BGH-supp. By Lemma 17 of BGH-supp
we have 
\begin{equation}
\tau_{\eta}(x^{\prime}\eta)=\tau_{0}(x^{\prime}\eta_{0})+(\eta-\eta_{0})(x-E[X|X^{\prime}\eta_{0}=x^{\prime}\eta_{0}])\tau_{0}^{\prime}(x^{\prime}\eta_{0})+o_{p}(\eta-\eta_{0}).\label{pf:derivative}
\end{equation}
For $II$, observe that 
\begin{eqnarray}
II & = & \int(x-E[X|x^{\prime}\hat{\eta}])\{u^{2}-\tau_{\hat{\eta}}(x^{\prime}\hat{\eta})\}dP(x,u)+\int(x-E[X|x^{\prime}\hat{\eta}])\hat{e}(w,u)dP(w,u)\nonumber \\
 & = & \left\{ \int(x-E[X|x^{\prime}\eta_{0}])(x-E[X|X^{\prime}\eta_{0}=x^{\prime}\eta_{0}])\tau_{0}^{\prime}(x^{\prime}\eta_{0})dP(x)\right\} (\hat{\eta}-\eta_{0})\nonumber \\
 &  & +\int(x-E[X|x^{\prime}\hat{\eta}])\hat{e}(w,u)dP(w,u)+o_{p}(\hat{\eta}-\eta_{0})\nonumber \\
 & = & \left\{ \int(x-E[X|x^{\prime}\eta_{0}])(x-E[X|x^{\prime}\eta_{0}])\tau_{0}^{\prime}(x^{\prime}\eta_{0})dP(x)\right\} (\hat{\eta}-\eta_{0})+O_{p}(n^{-1/2})+o_{p}(\hat{\eta}-\eta_{0})\nonumber \\
 & = & B(\hat{\eta}-\eta_{0})+O_{p}(n^{-1/2})+o_{p}(\hat{\eta}-\eta_{0}),\label{eq:score_2}
\end{eqnarray}
where the third equality follows from (\ref{pf:derivative}) and $(E[X|x^{\prime}\hat{\eta}]-E[X|x^{\prime}\eta_{0}])(\hat{\eta}-\eta_{0})=o_{p}(\hat{\eta}-\eta_{0})$,
the fourth equality follows from $\hat{\theta}_{\mathrm{OLS}}-\theta=O_{p}(n^{-1/2})$
and the definition of $B$ in Assumption M6.

Combining (\ref{pf:score_decompo}), (\ref{pf:score_1}), and (\ref{eq:score_2}),
we have 
\[
\hat{\eta}-\eta_{0}=B^{-}\int(x-E[X|x^{\prime}\eta_{0}])\{u^{2}-\tau_{0}(x^{\prime}\eta_{0})\}d(\mathbb{P}_{n}-P)(x,u)+O_{p}(n^{-1/2})+o_{p}(n^{-1/2}+(\hat{\eta}-\eta)),
\]
where $B^{-}$ is the Moore-Penrose inverse of $B$ (see p.17 of BGH
for more details). Therefore, we have $\hat{\eta}-\eta_{0}=O_{p}(n^{-1/2})$.
This result, combined with (\ref{pf:derivative}) and Assumptions
M1 and M2, implies $\tau_{\hat{\eta}}(a)-\tau_{0}(a)=O_{p}(n^{-1/2})$
and $\left\Vert \tau_{\hat{\eta}}-\tau_{0}\right\Vert _{2,P}=O_{p}(n^{-1/2})$.

\end{document}